\documentclass[10pt, twocolumn,secnumarabic,amssymb, nobibnotes, superscriptaddress, nofootinbib,aps,prd]{revtex4-2}
\usepackage[hidelinks,hypertexnames=false]{hyperref}
\usepackage{natbib}
\usepackage{etoolbox}
\DeclareRobustCommand{\Eprint}[2]{\href{#1}{\mbox{#2}}}
\let\BibliographyHref\href
\newcommand{\UnbrokenBibliographyHref}[2]{\BibliographyHref{#1}{\mbox{#2}}}
\AtBeginEnvironment{thebibliography}{\let\href\UnbrokenBibliographyHref}

\usepackage{amsmath}
\usepackage{adjustbox}
\usepackage{graphicx}
\usepackage{booktabs}       % professional-quality tables
\usepackage{longtable}

\usepackage{amsfonts}       % blackboard math symbols
\usepackage{nicefrac}       % compact symbols for 1/2, etc.
\usepackage{microtype}
\usepackage[capitalise]{cleveref}
\usepackage{xcolor}
\usepackage{tcolorbox}
\usepackage{multirow}
\usepackage{makecell}
\usepackage{array}
\usepackage{orcidlink}

\begin{document}

\author{Prashant Thakur~\orcidlink{0000-0003-4189-6176}}
\email{prashant@yonsei.ac.kr}
\affiliation{Department of Physics, Yonsei University, Seoul, 03722, South Korea}

\author{Ishfaq Ahmad Rather~\orcidlink{0000-0001-5930-7179}}
\email{rather@astro.uni-frankfurt.de}
\affiliation{Institut f\"{u}r Theoretische Physik, Goethe Universit\"{a}t,
Max-von-Laue-Str.~1, D-60438 Frankfurt am Main, Germany}

\author{Y. Lim~\orcidlink{0000-0001-9252-823X}} % Add the ORCID ID if available
\email{ylim@yonsei.ac.kr}
\affiliation{Department of Physics, Yonsei University, Seoul, 03722, South Korea}

\title{Dark matter in neutron stars: two-fluid \texorpdfstring{$f$}{f}-mode oscillations in full
general relativity}

\begin{abstract}
  A dark-matter core and a soft nuclear equation of state can produce
  similar mass--radius signatures, motivating a complementary probe
  through the $f$-mode spectrum. We compute quadrupolar $f$-mode
  frequencies and gravitational-wave damping times in full general
  relativity for cold, nonrotating neutron stars admixed with
  self-interacting bosonic dark matter (DM). Our two-fluid survey spans
  21 nuclear equations of state covering nucleonic, hyperonic,
  $\Delta$-admixed, and combined hyperon--$\Delta$ compositions at
  $F_{\rm DM}=1$--$20\%$. The main sequences use $\lambda=\pi$ and
  boson masses of $500$ and $100\,\mathrm{MeV}$ for the core and halo
  studies, respectively. Independent normal-matter (NM) and DM
  displacements couple through the metric, yielding NM-led and DM-led
  $f$-mode branches whose radiative damping is determined by outgoing-wave
  boundary conditions. At $1.4\,M_\odot$, the core DM-led damping time
  decreases from order $10^3\,\mathrm{s}$ at $F_{\rm DM}=1\%$ to order
  $1\,\mathrm{s}$ at $20\%$, approximately following an empirical
  $F_{\rm DM}^{-2}$ dependence, while NM-led damping times remain within
  $0.12$--$0.33\,\mathrm{s}$. Halo DM-led damping instead increases with
  dark fraction: at the same stellar mass and $F_{\rm DM}=20\%$,
  dark radii approach $88\,\mathrm{km}$, frequencies fall to
  $0.14\,\mathrm{kHz}$, and damping times reach approximately
  $450\,\mathrm{s}$. Selected high-mass core sequences exhibit strong
  DM-led damping-time upturns not reproduced by the leading
  mass-quadrupole diagnostic. Representative core calculations show
  greater DM-led sensitivity to the bosonic self-coupling and retain
  both branches at $\ell=3$. These results show how dark-core and
  dark-halo configurations affect the oscillation spectrum and radiative
  damping. Under the adopted nonrotating, circular-binary prescriptions,
  no evaluated configuration satisfies all geometric and perturbative
  conditions for a controlled linear tidal resonance.
  \end{abstract}

\maketitle

\section{Introduction}

  The binary-neutron-star event GW170817 provided the first direct
  gravitational-wave (GW) constraint on neutron-star tidal deformability, with
  the initial low-spin analysis yielding $\tilde{\Lambda}\leq800$ at the $90\%$
  credible level~\cite{LIGOScientific:2017vwq}. Complementary mass--radius
  posteriors have been obtained from the Neutron Star Interior Composition
  Explorer (NICER) pulse-profile analyses of PSR~J0030$+$0451
  \cite{Miller:2019cac,Riley:2019yda}, PSR~J0740$+$6620
  \cite{Miller:2021qha,Riley:2021pdl}, PSR~J0437$-$4715
  \cite{Choudhury:2024xbk}, and PSR~J0614$-$3329
  \cite{Mauviard:2025dmd}. Most recently, a joint NICER, XMM--Newton, and
  Chandra analysis of PSR~J1614$-$2230 reported
  $M=1.937^{+0.012}_{-0.013}\,M_\odot$ and
  $R_{\rm eq}=10.06^{+1.25}_{-0.87}\,\mathrm{km}$, with equally tailed
  $68\%$ credible intervals~\cite{Mauviard:2026gzc}. Together, these
  measurements disfavor equations of state (EOS) that are either too soft to
  support massive stars or too stiff to satisfy radius and tidal bounds, while
  leaving the high-density composition unresolved.  At the low-mass end, carbon-atmosphere X-ray spectral modeling of the central
  compact object in the supernova remnant HESS~J1731$-$347 has been used to
  infer an unusually light neutron star with $M\simeq0.77\,M_\odot$ and
  $R\simeq10.4\,\mathrm{km}$; these values, however, remain sensitive to the
  assumed atmosphere and source properties
  \cite{Doroshenko:2022nwp,Sagun:2023rzp}. Static mass--radius and tidal
  observables primarily constrain the equilibrium structure, whereas
  quasinormal-mode (QNM) asteroseismology probes its dynamical response. In
  particular, the fundamental ($f$) mode of an isolated neutron star typically
  lies at $1.5$--$3\,\mathrm{kHz}$ and has a gravitational-wave damping time of
  $0.1$--$2\,\mathrm{s}$; its frequency and damping time correlate with the
  stellar mean density and compactness through approximately EOS-insensitive
  relations~\cite{Andersson:1997rn, Lau:2009bu,Lioutas:2021jbl,Rather:2024nry}.

  Neutron stars are promising laboratories for dark matter (DM): their deep
  gravitational potentials focus ambient particles, while scattering with
  stellar constituents can dissipate sufficient kinetic energy for capture and
  subsequent accumulation~\cite{Bell:2020obw,Bramante:2023djs, Kouvaris:2010vv, deLavallaz:2010wp, Bertone:2007ae, Goldman:1989nd}. For asymmetric
  bosonic dark matter, a conserved dark charge supports a particle--antiparticle
  asymmetry that suppresses self-annihilation, allowing an accumulated component
  to persist over stellar timescales~\cite{Petraki:2013wwa, Zurek:2013wia}. We model this component as a massive
  complex scalar field with a repulsive quartic self-interaction,
  $V_{\rm int}(\phi)=\lambda|\phi|^4/4$, whose zero-temperature mean-field
  condensate obeys an analytical barotropic EOS~\cite{Karkevandi:2021ygv, Colpi:1986ye, Pitz:2023ejc}. Its
  characteristic transition density, $\epsilon_{\rm tr}=m_\chi^4/(3\lambda)$,
  implies that a smaller $m_\chi$ at fixed $\lambda$ favors a more extended dark
  component. The actual core--halo classification also depends on the
  dark-matter fraction $F_{\rm DM}$, the normal-matter EOS, and the relative
  central enthalpies of the two fluids~\cite{Miao:2022rqj, Giangrandi:2022wht}. Both geometries have
  been studied through their equilibrium mass--radius and tidal properties
  \cite{Karkevandi:2021ygv,Nelson:2018xtr,Miao:2022rqj, Sandin:2008db, Zhang:2025rur}; their nonradial spectra
  require a two-fluid perturbation treatment that solves the independent fluid
  displacements together with the spacetime perturbations in general relativity
  (GR).

  Allowing the normal-matter (NM) and DM components to move independently
  changes the polar-mode spectrum qualitatively. Each eigenmode contains both
  fluid displacements coupled through the common metric perturbations, producing
  two principal $f$-like families whose local character may be NM-led, DM-led,
  or mixed. The additional relative-fluid family has no counterpart in an
  effective one-displacement treatment. Earlier full-GR calculations of DM
  effects on the $f$ mode employed effective single-fluid descriptions
  \cite{Shirke:2024ymc,Thakur:2025zhi}, whereas two-fluid Cowling studies
  identified component-associated modes in core and halo configurations but
  could not determine their gravitational-wave damping times
  \cite{Sotani:2025hzb,Sotani:2025lzy}. Kumar et al.\ subsequently formulated
  the full-GR polar problem for independently conserved, gravitationally coupled
  fluids and computed representative real-frequency spectra for mirror-DM stars,
  without extracting radiative damping times~\cite{Kumar:2026tgi}. Zheng et al.\
  later imposed an outgoing-wave condition and obtained complex frequencies,
  damping times, and core-to-halo mode evolution for self-interacting fermionic
  DM using one microscopic normal-matter EOS~\cite{Zheng:2026fwf}.

  Here we perform a systematic full-GR complex-mode survey for self-interacting
  bosonic DM using 21 cold, beta-equilibrated normal-matter EOS spanning
  nucleonic, hyperonic, $\Delta$-admixed, and combined hyperon--$\Delta$
  compositions. We consider
  $F_{\rm DM}=M_{\rm DM}/(M_{\rm NM}+M_{\rm DM})=1\%,5\%,10\%$, and $20\%$,
  and construct both dark-core and dark-halo sequences. For each configuration,
  we solve the coupled polar perturbation equations for the two independent
  fluid displacements and their shared metric perturbations. Matching the
  quadrupolar ($\ell=2$) solution to a purely outgoing Zerilli wave yields the
  complex eigenfrequency and hence both the mode frequency and its radiative
  damping time. We track the two $f$-like families by spectral continuity and
  classify their local character using the node structure and relative
  amplitudes of the two fluid displacements, supplemented in halo configurations
  by frequency proximity to the corresponding single-fluid NM mode. We
  additionally calculate the octupolar ($\ell=3$) modes for a representative EOS,
  examine the influence of the bosonic self-coupling within the core
  configuration, and compare the equilibrium mass--radius sequences with
  published observational regions inferred from GW170817, NICER, and
  HESS~J1731$-$347.

  Across the EOS ensemble, the NM-led family remains continuously connected to
  the ordinary single-fluid $f$ mode, whereas the additional DM-sensitive family
  responds strongly to the dark fraction, spatial distribution, and
  self-interaction. Extended halos reach radii approaching $90\,\mathrm{km}$ at
  $F_{\rm DM}=20\%$, shifting the DM-led oscillation into the sub-kilohertz
  regime and substantially increasing its radiative damping time, while selected
  core configurations become weakly radiating. The broad EOS coverage separates
  these geometry-driven effects from variations associated with the
  normal-matter composition. Finally, for the nonrotating circular binaries
  evaluated here, none of the isolated-star modes satisfies all necessary
  conditions for a clean linear resonance before contact or Roche overflow. The
  tidal analysis therefore determines the accessibility of this excitation
  channel rather than establishing detector-level detectability.

  The paper is organized as follows. The normal- and dark-matter EOS are
  described in \cref{subsec:nuclear_eos,sec:dm_eos}. The full-GR two-fluid
  perturbation framework is presented in \cref{sec:twofluid_modes}. The numerical
  implementation and validation tests are described next, followed by the
  tidal-dephasing formalism in \cref{sec:dephasing}. The core and halo results
  are discussed in \cref{sec:core_results,sec:halo_results}, and the
  accessibility of tidal resonances is examined in \cref{sec:res_access}.
  \Cref{sec:discussion} presents the physical interpretation, observational
  prospects, comparison with previous work, and limitations. The detailed tidal
  derivation, representative eigenfunctions, dark-matter self-interaction scan,
  higher-multipole results, and numerical diagnostics are collected in the
  Appendices. Finally, \cref{conc} summarizes the main findings.

%%%%%%%%%%%%%%%%%%%%%%%%%%%%%%%%%%%%%%%%%%%%%%%%%%%%%%%%%%%%%%%%%%%%%%%%%%%%%%%%%%%%%

\section{Methodology}

\subsection{Nuclear-matter equations of state}
  \label{subsec:nuclear_eos}

For the normal-matter sector we use 21 cold, beta-equilibrated equations of state, chosen to sample the two axes along which dense matter is least well determined: its composition, and its stiffness at supra-saturation density.

Where present, the model-name suffix identifies the composition: $N$ denotes
  purely nucleonic $npe\mu$ matter, $Y$ the inclusion of the hyperon octet, and
  $D$ the inclusion of the $\Delta(1232)$ isobars; $NYD$ contains both exotic
  species. For the parameterizations of Set~3, the label also encodes
the two nuclear-matter parameters that are varied: the nucleon effective mass $m^{*}/m_N$ at saturation and symmetry-energy slope $L$.
These parameters probe different aspects of the EOS: variations in
  $m^{*}/m_N$ predominantly affect the high-density stiffness and
  $M_{\max}$, whereas variations in $L$ predominantly affect the pressure near
  saturation and $R_{1.4}$.

The ensemble is organized into three sets. \emph{Set~1} comprises five unified nucleonic models (EOS1, EOS2, EOS6, EOS11, and EOS20) selected from the nonlinear relativistic mean-field
  ensemble of Malik, Pais, and Provid\^encia~\cite{Malik:2024nva}, sampling a
  broad range of nucleonic stiffnesses. \emph{Set~2} combines three separately supplied DD2-based EOS with five
  models extracted from the CompOSE database
  \cite{Typel:2013rza,Oertel:2016bki}. The first group comprises DD2\_ND,
  containing $\Delta$ isobars; DD2\_NY, containing the hyperon octet; and
  DD2\_NYD, containing both hyperons and $\Delta$ isobars. The CompOSE group
  comprises the hyperonic DD2Y model~\cite{Marques:2017zju}; the
  density-dependent hyperonic DDHdY4 model
  \cite{Gaitanos:2003zg,Oertel:2014qza}; the nonlinear relativistic mean-field
  GM1Y5 model with hyperons
  \cite{Glendenning:1991es,Oertel:2014qza}; the chiral mean-field CMF-crust
  model~\cite{Dexheimer:2008ax,Schurhoff:2010ph,Dexheimer:2015qha,
  Gulminelli:2015csa}; and the purely nucleonic, nonrelativistic Skyrme model
  SLy4~\cite{Chabanat:1997un,Douchin:2001sv}.

\emph{Set~3} comprises eight nonlinear relativistic mean-field (RMF)
  parameterizations based on the parameter-variation scheme of
  Ref.~\cite{Hornick:2018kfi}, with the symmetry energy at saturation fixed at
  $J=30~\mathrm{MeV}$. The four hyperonic models use
  $m^{*}/m_N=0.55$ or $0.60$ with $L=50$ or $60~\mathrm{MeV}$, whereas the
  four nucleonic models use $m^{*}/m_N=0.70$ or $0.75$ with $L=40$ or
  $50~\mathrm{MeV}$. Thus, within each fixed composition and effective mass,
  the models form pairs differing only in $L$; effective mass and composition
  remain correlated across the set and cannot be separated independently.

Hyperons and $\Delta$ isobars can soften the EOS by redistributing the
  baryonic degrees of freedom, but their onset and effects on $M_{\max}$ and
  $R_{1.4}$ depend on the meson--baryon couplings
  \cite{Oertel:2016bki,Lopes:2022vjx}. Since composition and dark matter both
  modify stellar compactness and the $f$-mode spectrum, the EOS survey is needed
  to distinguish their effects. The model-by-model results are reported in
  \cref{tab:core_config_full_gr}.

%%%%%%%%%%%%%%%%%%%%%%%%%%%%%%%%%%%%%%%%%%%%%%%%%%%%%%%%%%%%%%%%

 \subsection{Bosonic dark-matter equation of state}
  \label{sec:dm_eos}

DM in neutron stars has been studied through capture and
  accumulation constraints~\cite{Goldman:1989nd}, effective single-fluid
  descriptions of fermionic admixture~\cite{Panotopoulos:2017idn,
  Sen:2021wev, Thakur:2025oyn, Thakur:2025zhi}, and two-fluid equilibrium
  models~\cite{Xiang:2013xwa, Mariani:2023wtv, Thakur:2023aqm}.
  Here we consider bosonic DM throughout.
  We model the dark component as cold asymmetric bosonic matter described by
  a massive complex scalar field with a conserved charge and a repulsive
  quartic self-interaction,
  $V_{\rm int}(\phi)=\lambda|\phi|^{4}/4$.  At zero temperature the bosons
  are assumed to form a Bose--Einstein condensate, for which the mean-field
  equation of state is~\cite{Karkevandi:2021ygv}
  \begin{equation}
  p_{\chi}(\epsilon_{\chi})
  =
  \frac{m_{\chi}^{4}}{9\lambda}
  \left[
  \sqrt{1+\frac{3\lambda\epsilon_{\chi}}{m_{\chi}^{4}}}
  -1
  \right]^{2}.
  \label{eq:bosonic_dm_eos}
  \end{equation}
  Here $m_{\chi}$ is the boson mass, $\lambda>0$ is the dimensionless
  repulsive self-coupling, and $\epsilon_{\chi}$ and $p_{\chi}$ denote the
  total dark-matter energy density and pressure, respectively.  \cref{eq:bosonic_dm_eos} is written in natural units,
  $\hbar=c=1$, and applies in the strong-coupling mean-field regime
  \begin{equation}
  \lambda\gg4\pi
  \left(\frac{m_{\chi}}{M_{\rm Pl}}\right)^{2},
  \end{equation}
  which is amply satisfied for the parameter values considered here.

  The characteristic energy density separating the two asymptotic regimes is
  \begin{equation}
  \epsilon_{\rm tr}
  =
  \frac{m_{\chi}^{4}}{3\lambda}.
  \end{equation}
  Consequently,
  \begin{equation}
  p_{\chi}\simeq
  \begin{cases}
  \dfrac{\lambda}{4m_{\chi}^{4}}\epsilon_{\chi}^{2},
  & \epsilon_{\chi}\ll\epsilon_{\rm tr},\\[6pt]
  \dfrac{1}{3}\epsilon_{\chi},
  & \epsilon_{\chi}\gg\epsilon_{\rm tr}.
  \end{cases}
  \label{eq:bosonic_dm_limits}
  \end{equation}
  Thus, at fixed $\lambda$, decreasing $m_{\chi}$ shifts the transition to
  lower density and produces a more extended dark component, whereas larger
  $m_{\chi}$ favors a compact dark core.

  Unless stated otherwise, we adopt the representative coupling
  $\lambda=\pi$.  We use $m_{\chi}=500~{\rm MeV}$ for the compact
  DM-core sequences and $m_{\chi}=100~{\rm MeV}$ for the extended
  DM-halo sequences.  These choices sample the two qualitatively distinct
  spatial configurations of the same microscopic model.  The quartic
  interaction acts only among dark-matter particles; the normal and dark
  fluids remain coupled exclusively through the common spacetime metric.
  For the numerical tables, \cref{eq:bosonic_dm_eos} is evaluated by
  converting
  $\epsilon_{\chi}[{\rm MeV\,fm^{-3}}]$ to natural units through
  $\epsilon_{\chi}^{\rm nat}=(\hbar c)^{3}\epsilon_{\chi}$, with
  $\hbar c=197.3269804~{\rm MeV\,fm}$, and converting the resulting pressure
  back to ${\rm MeV\,fm^{-3}}$.

We generalize the Tolman-Oppenheimer-Volkoff (TOV) equations to two fluids that couple only through the metric, the standard framework for dark-matter-admixed stars \cite{Grippa:2024ach, Leung:2022wcf}. Because the two fluids interact only gravitationally, the usual one-fluid stability analysis does not apply. 
The turning-point theorem applies to a one-parameter equilibrium sequence, whereas the two-fluid configurations form a two-parameter family labelled by the two central pressures; along a slice of fixed $F_\text{DM}$ the sign change of $dM/d\varepsilon_c$ therefore need not coincide with the onset of instability.
Instead, we use the radial density oscillations for both dark matter and normal matter fluids as discussed in  Ref.~\cite{Hippert:2022snq} and have been used in several studies \cite{Biesdorf:2024dor, Pitz:2024xvh, Barbat:2024yvi, Rather:2026aed}. A configuration is retained only where the fundamental radial eigenfrequency satisfies $\omega^2_0 = 0$. 

%%%%%%%%%%%%%%%%%%%%%%%%%%%%%%%%%%%%%%%%%%%%%%%%%%%%%%%%%%%%%%%

\subsection{Full-GR two-fluid polar modes}
\label{sec:twofluid_modes}

We compute the quadrupolar mode frequencies and gravitational-wave damping
times using the fully relativistic polar-perturbation formalism of
Ref.~\cite{Kumar:2026tgi}. The normal-matter and dark-matter components,
$X\in\{{\rm N},{\rm D}\}$, are modeled as independently conserved barotropic
perfect fluids,
\begin{equation}
\begin{aligned}
&G_{\mu\nu}
 =8\pi\sum_X T^X_{\mu\nu},\\
&\nabla_\mu T_X^{\mu\nu}
 =0.
\end{aligned}
\end{equation}
The NM and DM fluids are separately conserved and interact only through the
common spacetime~\cite{Andersson:2006nr}; no direct non-gravitational
interaction or particle conversion between them is included. This gravity-only system belongs to the broader class of relativistic
  multifluid models, but it is not generally obtained from the standard
  superfluid formalism merely by setting the entrainment to zero, because the
  two constituents can have different equations of state and distinct radial
  supports~\cite{Andersson:2006nr,Kumar:2026tgi}. Thermodynamically consistent two-fluid equilibrium models with
  Higgs-mediated DM--nucleon interactions have also been
  developed~\cite{Hajkarim:2024ecp}.
  Accordingly, the NM-led and
  DM-led labels identify the component that dominates a given eigenfunction;
  they do not represent a co-moving/counter-moving decomposition. In contrast to
the mirror-matter examples of Ref.~\cite{Kumar:2026tgi}, we allow
$p_{\rm N}(\epsilon_{\rm N})\neq p_{\rm D}(\epsilon_{\rm D})$ and consider both
DM-core and DM-halo configurations.

Using $G=c=1$, the equilibrium metric is written as
  \begin{equation}
  \begin{aligned}
  ds^2={}&-e^{2\Phi(r)}dt^2
          +e^{2\Lambda(r)}dr^2+r^2d\Omega^2,\\
  e^{-2\Lambda(r)}={}&1-\frac{2m(r)}{r},
  \end{aligned}
  \end{equation}
  where $m(r)=\sum_Xm_X(r)$ is the total mass function. We retain the metric
  potentials $\Phi$ and $\Lambda$ throughout, reserving $\lambda$ exclusively
  for the dark-matter self-coupling.

  In Regge--Wheeler gauge
  \cite{Regge:1957td,Zerilli:1970se,Lindblom:1983ps}, an even-parity
  perturbation with $\ell\geq2$ and time dependence $e^{i\omega t}$ is described
  by the shared metric amplitudes $(H_0,H_1,K)$, with $H_2=H_0$, and by the
  independent displacement amplitudes $(W_X,V_X)$ of each fluid
  \cite{Lindblom:1983ps,Kumar:2026tgi}. For a fixed $(\ell,m)$ mode, we define
  the radial Lagrangian-pressure amplitude ${\cal P}_X$ through
  \begin{equation}
  \Delta_Xp_X
  =
  -e^{-\Phi}r^\ell{\cal P}_X(r)Y_{\ell m}e^{i\omega t}.
  \end{equation}
  The two-fluid problem then consists of six first-order equations,
  \begin{equation}
  \begin{aligned}
  \mathbf{y}'
   &=
   \mathbf{A}_{\ell}(r,\omega)\mathbf{y},\\
  \mathbf{y}
   &=
   \bigl(
   H_1,K,W_{\rm N},{\cal P}_{\rm N},
   W_{\rm D},{\cal P}_{\rm D}
   \bigr)^{\mathsf T},
  \end{aligned}
  \label{eq:twofluid_state}
  \end{equation}
  supplemented by algebraic relations for $H_0$ and $V_X$. We use the complete
  system given in Eqs.~(23)--(29) of Ref.~\cite{Kumar:2026tgi}, without invoking
  the Cowling approximation.

  The gravitational coupling is manifest in the fluid-summed source terms
  $\sum_X(\epsilon_X+p_X)W_X$ in $K'$ and
  $\sum_X(\epsilon_X+p_X)V_X$ in $H_1'$, together with the cross-fluid
  combination
  $(\epsilon_X+p_X)\sum_Y(\epsilon_Y+p_Y)W_Y$ in ${\cal P}'_X$.
  Thus, each component retains its own displacement field while both fluids
  couple through the common background and metric perturbations that they
  jointly source. The central regularity conditions are imposed simultaneously
  on both fluids as in Ref.~\cite{Kumar:2026tgi}.

\begin{comment}
Its shared gravitational sources are explicit, for
example,
\begin{align}
K'={}&
 -\left(\frac{\ell+1}{r}-\Phi'\right)K
 +\frac{\ell(\ell+1)}{2r}H_1
 \nonumber\\
 &+\frac{H_0}{r}
 +\frac{8\pi e^\Lambda}{r}
  \sum_X(\epsilon_X+p_X)W_X ,
\\
H_1'={}&
 \left[
  \Lambda'-\Phi'-\frac{\ell+1}{r}
 \right]H_1
 +\frac{e^{2\Lambda}}{r}(H_0+K)
 \nonumber\\
 &+\frac{16\pi e^{2\Lambda}}{r}
  \sum_X(\epsilon_X+p_X)V_X ,
\end{align}
while the pressure equation contains the cross-fluid term
\begin{equation}
\begin{aligned}
{\cal P}'_X\supset{}&
 -\frac{4\pi e^{\Lambda+\Phi}}{r}
  (\epsilon_X+p_X)\\[-2pt]
 &{}\times
  \sum_Y(\epsilon_Y+p_Y)W_Y .
\end{aligned}
\label{eq:twofluid_cross_source}
\end{equation}

Thus, the NM and DM fluids have distinct displacement fields,
while both contribute to and are coupled through the same metric perturbation.
Regularity at the center requires
\begin{align}
H_0(0)=&K(0),&\\
W_X(0)=&-\ell V_X(0),
\\
H_1(0)={}&
 \frac{2K(0)}{\ell+1}
 -\frac{16\pi}{\ell(\ell+1)}
 \nonumber\\[-2pt]
 &{}\times
 \sum_X[\epsilon_X(0)+p_X(0)]W_X(0),
\\
{\cal P}_X(0)={}&
 [\epsilon_X(0)+p_X(0)]e^{\Phi(0)}
 \nonumber\\[-2pt]
 &{}\times
 \Biggl[
 \Biggl\{
 \frac{4\pi}{3}
 \sum_Y[\epsilon_Y(0)+3p_Y(0)]
 \nonumber\\[-2pt]
 &\hspace{3.3em}
 -\frac{\omega^2e^{-2\Phi(0)}}{\ell}
 \Biggr\}W_X(0)
 -\frac{K(0)}{2}
 \Biggr].
\label{eq:twofluid_center}
\end{align}
\end{comment}
Each component terminates at its own surface $R_X$, where
  \begin{equation}
  \begin{aligned}
  p_X(R_X)&=0,\\
  \Delta_Xp_X(R_X)=0
  &\quad\Longleftrightarrow\quad
  {\cal P}_X(R_X)=0.
  \end{aligned}
  \label{eq:twofluid_surface}
  \end{equation}
  We denote the smaller and larger component radii by
  \begin{equation}
  R_I=\min(R_{\rm N},R_{\rm D}),
  \qquad
  R_O=R_{\rm out}=\max(R_{\rm N},R_{\rm D}),
  \end{equation}
  respectively. At the inner surface $R_I$, the metric amplitudes and the
  surviving outer-fluid variables $(H_1,K,W_O,{\cal P}_O)$ remain continuous.
  For $r>R_I$, the terminated component is removed by setting
  $\epsilon_I=p_I=0$ and dropping $(W_I,{\cal P}_I)$ from the state vector.
  Consequently, in the region $R_I<r<R_O$, the perturbation equations involve
  only the surviving outer-fluid variables $(H_1,K,W_O,{\cal P}_O)$. The vacuum
  exterior begins at $R_{\rm out}=R_O$. The inner component is dark matter in a
  DM-core configuration and normal matter in a DM-halo configuration.

For each trial frequency $\omega$, we construct two independent solutions
  satisfying the central regularity conditions. In each solution, the remaining
  free central displacement amplitude is fixed by imposing the inner-surface
  condition ${\cal P}_I(R_I)=0$. A linear combination of these two solutions is
  then chosen to satisfy the outer-surface condition
  ${\cal P}_O(R_O)=0$. Because the perturbation equations and boundary
  conditions are linear, this outward-shooting construction imposes the same
  global interior boundary conditions as the two-sided matching procedure of
  Ref.~\cite{Kumar:2026tgi}.

  At $R_{\rm out}=R_O$, the interior metric perturbations are mapped to the
  gauge-invariant Zerilli function $Z$, which satisfies
  \begin{equation}
  \begin{aligned}
  \frac{d^2Z}{dr_*^2}
  +\left[\omega^2-V_\ell^{\rm Z}(r)\right]Z
  &=0,\\
  r_*
  &=r+2M\ln\left(\frac{r}{2M}-1\right).
  \end{aligned}
  \end{equation}
  Here $M=m(R_{\rm out})$ is the total gravitational mass and
  $V_\ell^{\rm Z}(r)$ is the standard Zerilli potential
  \cite{Zerilli:1970se}.

\begin{comment}
\begin{equation}
\begin{aligned}
V_\ell^{\rm Z}(r)&=
 \left(1-\frac{2M}{r}\right) \\
 & \frac{2n^2(n+1)r^3+6n^2Mr^2+18nM^2r+18M^3}
      {r^3\,(nr+3M)^2},
\end{aligned}
\end{equation}
with $n=(\ell-1)(\ell+2)/2$.
\end{comment}
With the adopted $e^{i\omega t}$ convention, the asymptotic solution has the
  form
  \begin{equation}
  Z\longrightarrow
  A_{\rm out}e^{-i\omega r_*}
  +
  A_{\rm in}e^{+i\omega r_*},
  \qquad
  r_*\rightarrow\infty.
  \end{equation}
  A quasinormal mode satisfies the purely outgoing condition
  \begin{equation}
  A_{\rm in}(\omega_\alpha)=0.
  \label{eq:twofluid_qnm}
  \end{equation}
  For an eigenfrequency expressed in inverse-length units, the oscillation
  frequency and radiative damping time are
  \begin{equation}
  f_\alpha
  =
  \frac{c\,\operatorname{Re}\omega_\alpha}{2\pi},
  \qquad
  \tau_\alpha
  =
  \frac{1}{c\,\operatorname{Im}\omega_\alpha},
  \label{eq:mode_frequency_damping}
  \end{equation}
  where $\operatorname{Im}\omega_\alpha>0$ corresponds to decay for the adopted
  time convention.

  Dense real-frequency scans are first used to locate and track the mode
  branches. Each real-frequency root then seeds a complex-frequency solution of
  the complete two-fluid interior and outgoing Zerilli exterior problem. The
  reported damping times are obtained from these complex roots; extended
  precision is used for weakly radiating modes whose imaginary parts approach
  the double-precision floor.

  The quadrupolar ($\ell=2$) fundamental branches are tracked by spectral
  continuity and classified using the node structure and relative amplitudes of
  the two fluid displacements. The NM-led branch is additionally identified by
  its continuous connection, or frequency proximity in halo configurations, to
  the corresponding single-fluid normal-matter $f$ mode. The DM-led branch has
  a nodeless dominant dark-matter displacement and no single-fluid counterpart.

%%%%%%%%%%%%%%%%%%%%%%%%%%%%%%%%%%%%%%%%%%%%%%%%%%%%%%%%%%%
%%%%%%%%%%%%%%%%%%%%%
\subsection{Numerical implementation and validation}

%\IR{Can we move this to appendix? It breaks the flow of the paper here and also not very important to discuss it here.}
  The two-fluid TOV equations are integrated with a fourth-order Runge--Kutta
  scheme using a default radial spacing $\Delta r=0.5\,\mathrm{m}$. For each
  trial complex frequency, the regular interior basis solutions are evolved
  outward, the two fluid-surface conditions are imposed by affine shooting and
  linear combination, and the resulting solution is matched to the outgoing
  Zerilli exterior. The complex eigenfrequency is obtained by solving
  $A_{\rm in}(\omega_\alpha)=0$ with a complex Newton iteration. The default
  calculation uses double-precision complex arithmetic; weakly radiating modes
  are recomputed with compiled long-double and IEEE binary128 implementations,
  while the most cancellation-sensitive poles are refined with
  $40$--$60$-decimal-digit arbitrary precision.

  Convergence is tested against the interior radial spacing, exterior step and
  matching radius, initial root seed, and arithmetic precision. In the
  single-fluid limit, the mass agrees at the quoted precision, while the radius
  and NM-led frequency agree within $0.013\%$ and $0.001\%$, respectively. At
  real frequency, the complex interior reproduces the original implementation
  to $10^{-15}$ for the NM-led mode and $3\times10^{-12}$ for the DM-led mode.
  For the $157$ halo targets with certified DM-led complex modes, independent
  long-double and binary128 implementations differ by at most $0.0025\%$ in
  frequency and $0.129\%$ in damping time. The largest final spatial-resolution
  change in the halo calculation is $0.253\%$ in the damping time, while
  halving the exterior step and doubling the matching radius changes it by at
  most $0.023\%$.

%%%%%%%%%%%%%%%%%%%%%%%%%%%%%%%%%%%%%%%%%%%%%%%%%%%%%%%%%%%%
\subsection{Resonant tidal dephasing}
  \label{sec:dephasing}

  We combine the full-GR stellar response with the leading stationary-phase,
  quadrupolar tidal-resonance reduction
  \cite{Lai:1993di, Kokkotas:1995xe, 1994ApJ...426..688R}. As an independent leading-multipole
  diagnostic, we define
  \begin{equation}
  \begin{aligned}
  J_\alpha
  &=
  \sum_X\int_0^{R_X}
  (\epsilon_X+p_X)e^{\Lambda-\Phi}
  \left(|W_X|^2+6|V_X|^2\right)r^4\,dr,\\
  Q_\alpha
  &=
  \int
  (\delta\rho_{\rm N}+\delta\rho_{\rm D})
  r^2Y_{2m}^{*}\,d^3x .
  \end{aligned}
  \label{eq:overlap_diagnostics}
  \end{equation}
  Here $\delta\rho_X=\delta\epsilon_X/c^2$, and the relative phase of the two
  fluid perturbations is retained in $Q_\alpha$. The quantity $J_\alpha$ is a
  fluid inertia diagnostic rather than an exact GR quasinormal-mode norm. This dynamical-resonance diagnostic is complementary to static-tide DM-admixed Love-number calculations~\cite{Dengler:2021qcq, Diedrichs:2023trk}. In the
  leading mass-quadrupole reduction,
  \begin{equation}
  \left(\tau_\alpha^{\rm quad}\right)^{-1}
  =
  \frac{4\pi G}{75c^5}\,
  \omega_\alpha^4
  \frac{|Q_\alpha|^2}{J_\alpha}.
  \label{eq:duality}
  \end{equation}

  The fully relativistic stellar response is characterized instead by the
  flux-normalized Zerilli reflection coefficient. For an isolated, weakly
  damped polar QNM,
  \begin{equation}
  \begin{aligned}
  \left|
  \operatorname*{Res}_{\omega_{\rm QNM}}\mathcal R
  \right|
  &=
  \frac{2\mathcal B_\alpha}{\tau_\alpha},
  \\
  \mathcal B_\alpha
  &=
  1+
  O\!\left[
  (\omega_\alpha\tau_\alpha)^{-1},
  (\tau_\alpha\Delta\omega_\alpha)^{-1}
  \right],
  \end{aligned}
  \label{eq:gr_reciprocity}
  \end{equation}
  where $\tau_\alpha$ is the radiative amplitude-damping time
  \cite{Miao:2025utd}. Combining this response with the leading tidal source and
  circular point-mass chirp gives the per-star phase-advance magnitude
  \begin{equation}
  \boxed{
  \Delta\Phi_\alpha
  =
  \mathcal C_\alpha
  \frac{375}{131072\,\pi^5}
  \frac{c^{15}}
  {G^5M^3M'M_t f_\alpha^6\tau_\alpha}
  },
  \label{eq:crown}
  \end{equation}
  where
  \begin{equation}
  \begin{aligned}
  f_\alpha&=\frac{\omega_\alpha}{2\pi}=f_{\rm GW,res},\\
  M_t&=M+M',
  \\
  \mathcal C_\alpha
  &=
  \mathcal B_\alpha^2
  \kappa_{\rm tide}\kappa_{\rm orb}.
  \end{aligned}
  \end{equation}
  Here $M$ and $M'$ are the resonant-star and companion masses. At the order
  used here, $\kappa_{\rm tide}=\kappa_{\rm orb}=1$; retaining these factors
  makes explicit that the tidal-source matching and orbital reduction are
  leading-order approximations. The derivation is given in
  Appendix~\ref{app:tidal}.

  Equation~\eqref{eq:crown} is used only when the resonance satisfies
  \begin{equation}
  \begin{aligned}
  t_{\rm res}
  &=
  \left(\frac{2}{m\dot\Omega_{\rm res}}\right)^{1/2}
  =
  \dot\Omega_{\rm res}^{-1/2},
  \qquad m=2,\\
  \omega_\alpha t_{\rm res}&\gg1,
  \qquad
  t_{\rm res}\ll\tau_\alpha,
  \qquad
  \Delta\omega_\alpha t_{\rm res}\gg1,\\
  D_{\rm res}
  &=
  \left[
  \frac{GM_t}{(\pi f_\alpha)^2}
  \right]^{1/3}
  >
  R_{{\rm out},1}+R_{{\rm out},2},\\
  \Delta E_\alpha
  &\ll
  \min\!\left(
  E_{{\rm nonlinear},\alpha},
  \mathcal F_{\rm GW}t_{\rm res},
  |E_{\rm orb}|
  \right).
  \end{aligned}
  \label{eq:resonance_gates}
  \end{equation}
  Here
  $\Delta\omega_\alpha=\min_{\beta\neq\alpha}
  |\omega_\beta-\omega_\alpha|$ and
  $R_{\rm out}=\max(R_{\rm N},R_{\rm D})$. The nonlinear-energy threshold is
  configuration dependent; no quantitative phase is quoted unless all these
  conditions are satisfied.

  For extended halos, material contact is supplemented by a Roche-overflow
  condition~\cite{Paczynski:1971xb,Lai:1993pa,Eggleton:1983rx}. For an
  equal-mass binary, the inner Lagrange point lies at $D/2$, giving the necessary
  geometric condition
  \begin{equation}
  f_{L1}
  =
  \frac{1}{\pi}
  \left[
  \frac{GM_t}{(2R_{\rm DM})^3}
  \right]^{1/2}.
  \label{eq:fL1}
  \end{equation}
  With
  $\alpha_h=(2\pi f_\alpha)^2R_{\rm DM}^3/(GM)$, this gives
  $f_\alpha/f_{L1}=\sqrt{\alpha_h}$. For the full unequal-mass test, we use
  Eggleton’s volume-equivalent Roche radius,
  \begin{equation}
  \frac{R_L}{D}
  =
  \frac{0.49\,q^{2/3}}
  {0.6\,q^{2/3}+\ln(1+q^{1/3})},
  \qquad
  q=\frac{M}{M'},
  \label{eq:eggleton}
  \end{equation}
  and require $R_{\rm DM}\leq R_L$ at $D=D_{\rm res}$. This is a
  quasi-equilibrium accessibility criterion, not a dynamical calculation of
  stripping or mass transfer.

%%%%%%%%%%%%%%%%%%%%%%%%%%%%%%%%%%%%%%%%%%%%%%%%%%%%%%%%%%%%%%%%%%%

\section{Results}

%%%%%%%%%%%%%%%%%%%%%%%%%%%%%%%%%%%%%%%%%%%%%%%%%%%%%%%%%%%

\subsection{Dark-matter core configurations}

\label{sec:core_results}

\begin{figure*}[!tb]
  \centering
    \includegraphics[width=0.85\linewidth]{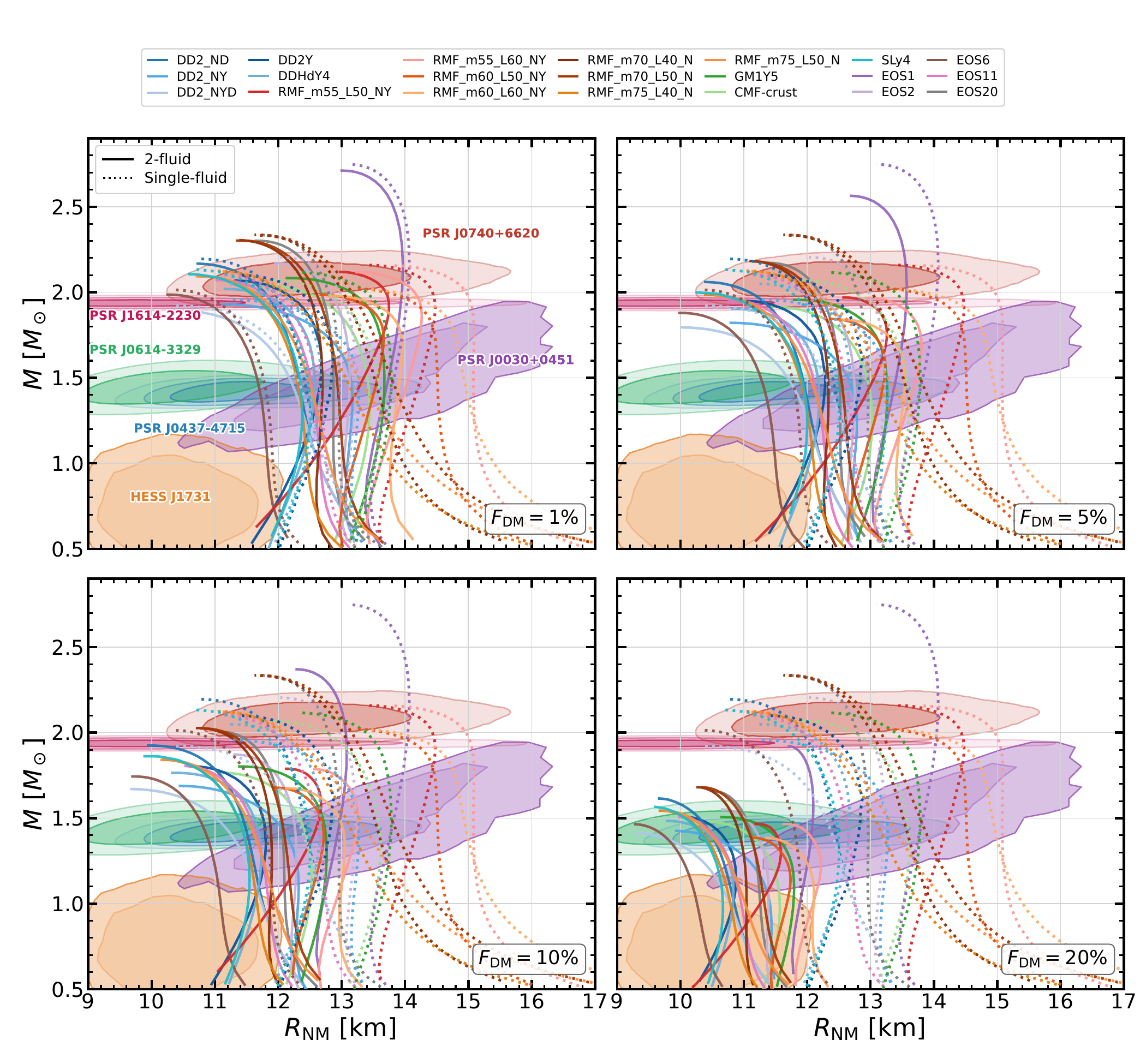}
  \caption{Mass--radius relations for the 21 nuclear EOS in the
  dark-matter-core configuration. The four panels correspond to
  $F_{\rm DM}=1\%,5\%,10\%$, and $20\%$. Solid curves show the two-fluid
  sequences, while dotted curves show the corresponding pure-normal-matter
  single-fluid references. Published mass--radius posterior regions
  inferred
  from GW170817 and NICER observations are overlaid for comparison
  \cite{LIGOScientific:2018cki,Riley:2019yda,Miller:2019cac,
  Riley:2021pdl,Miller:2021qha,Choudhury:2024xbk,Mauviard:2025dmd,
  Mauviard:2026gzc}. The model-dependent X-ray inference for the central
  compact object in HESS~J1731$-$347 is also shown
  \cite{Doroshenko:2022nwp}.}

  \label{mass_radius}
\end{figure*}

\cref{mass_radius} compares the mass--radius relations of the 21
nuclear EoS for dark-matter-core configurations with
$F_{\rm DM}=1\%, 5\%, 10\%$, and $20\%$. Solid curves denote the two-fluid stars, dotted curves the corresponding pure normal-matter sequences; since $R_{\rm DM}<R_{\rm NM}$, the plotted radius is the outer normal-matter radius $R_{\rm NM}$. Within the adopted strong-coupling mean-field model, the condensed bosons are
  supported by the repulsive quartic interaction rather than by Fermi
  degeneracy. Since the low-density limit of
  \cref{eq:bosonic_dm_eos} gives
  $p_\chi\simeq\lambda\epsilon_\chi^2/(4m_\chi^4)$, the
  $m_\chi=500\,\mathrm{MeV}$ component is relatively compressible at fixed
  $\lambda$ and remains concentrated inside the normal-matter star
  \cite{Karkevandi:2021ygv,Karkevandi:2024vov}. The two fluids
exchange no pressure directly: each obeys its own hydrostatic balance while responding to the gravitational field sourced by their combined energy density. Increasing $F_{\rm DM}$ places a larger share of the total mass in the inner dark component, which deepens the gravitational field the normal fluid sits in without supporting that fluid directly against compression. Its pressure then falls more rapidly with radius, moving the surface inward and systematically
reducing $R_{\rm NM}$ across the four panels; for this dark-matter EoS, the increased central concentration also lowers the maximum mass of each sequence, a trend specific to the adopted parameters and the core regime rather than a universal consequence of dark-matter admixture. At $F_{\rm DM}=1\%$, the reduction in $M_{\max}$ is small throughout the
  ensemble, but the radius response is strongly EOS-dependent: the Set~3 models
  contract by $\Delta R_{1.4}=0.85$--$1.10\,\mathrm{km}$, compared with at most
  $0.18\,\mathrm{km}$ for the Set~1 and Set~2 models. Both the maximum-mass
  reduction and the radius contraction strengthen as $F_{\rm DM}$ increases.
  For EOS1, which has the largest pure-normal-matter maximum mass,
  $M_{\max}$ decreases from $2.748$ to $1.920\,M_\odot$, while $R_{1.4}$
  decreases from $13.77$ to $12.07\,\mathrm{km}$ between the pure-normal-matter
  and $F_{\rm DM}=20\%$ configurations.

\begin{figure*}[!tb]
    \centering
    \includegraphics[width=0.85\linewidth]{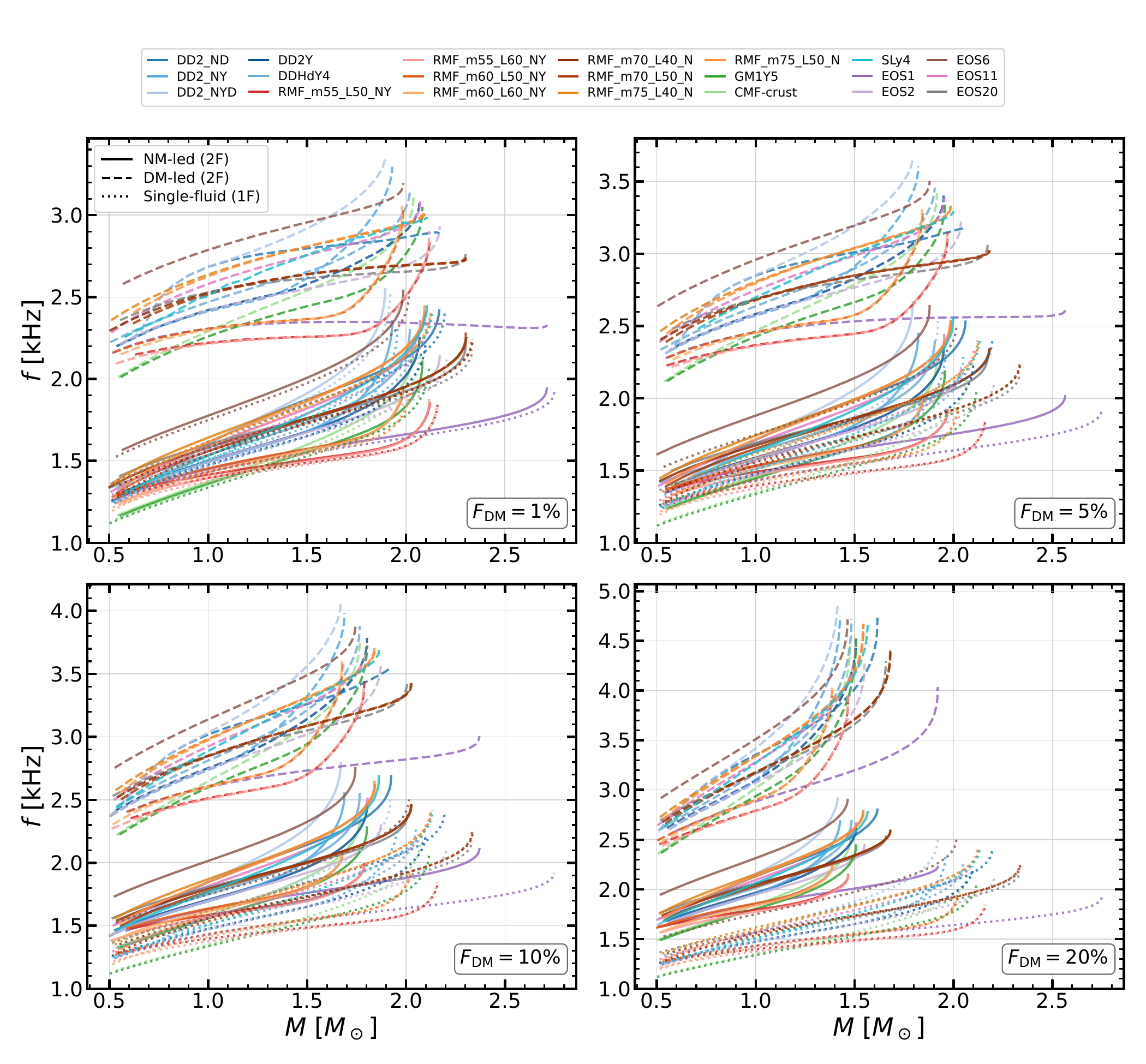}
  \caption{Full general-relativistic two-fluid $f$-mode frequencies as functions
  of gravitational mass for dark-matter-core configurations. The four panels
  correspond to $F_{\rm DM}=1\%,5\%,10\%$, and $20\%$. Colors identify the
  nuclear EOS listed in the legend. Solid and dashed curves denote the NM-led
  and DM-led two-fluid branches, respectively, while dotted curves show the
  corresponding pure-normal-matter single-fluid reference sequences.}

  \label{freq_mass}
\end{figure*}

The response nevertheless retains a clear dependence on the normal-matter
  composition, reflecting the pressure at different densities: the
  intermediate-density behavior primarily controls the radius, whereas
  sufficiently high-density pressure is required to retain a large maximum mass
  once the core is introduced. At $F_{\rm DM}=5\%$, every hyperon-bearing EOS
  in the ensemble falls below the nominal $2\,M_\odot$ threshold, whereas the
  stiffest nucleonic models and the $\Delta$-admixed DD2\_ND remain above it.
  Within this ensemble, EOS containing hyperons lose support for a
  $2\,M_\odot$ configuration at smaller dark-matter fractions because their
  high-density pressure is already reduced before the additional compression
  produced by the dark core. The controlled RMF pairs show that varying the
  symmetry-energy slope $L$ mainly changes $R_{1.4}$ while leaving $M_{\max}$
  nearly fixed, with a maximum-mass separation below $0.005\,M_\odot$. Their
  $R_{1.4}$ separation decreases as the dark core grows, indicating that the
  common compression induced by the dark component progressively dominates
  over the more modest radius difference produced by varying $L$. Moderate dark
  fractions can therefore move large-radius sequences toward the NICER and
  GW170817 credible regions, while the $20\%$ configurations show the greatest
  overlap with the HESS~J1731$-$347 region but all remain below
  $2\,M_\odot$. Within the adopted model, this indicates a tension between
  producing very compact low-mass stars and retaining massive configurations,
  rather than a universal constraint on the dark-matter fraction.

\cref{freq_mass} presents the full-GR quadrupolar $f$-mode frequencies of the dark-matter-core configurations of \cref{mass_radius}. Solid and dashed curves denote the normal-matter-led and dark-matter-led two-fluid branches, and dotted curves the pure normal-matter single-fluid reference. Two branches appear because each fluid carries its own displacement field while both source and respond to the same metric perturbation~\cite{Kumar:2026tgi, Zheng:2026fwf}: the NM-led branch is dominated by motion of the outer normal fluid and continuously approaches the single-fluid $f$-mode as $F_{\rm DM}\rightarrow0$, whereas the DM-led branch is carried by the inner dark component and has no counterpart when the dark fluid is removed. The higher DM-led frequency is set by the compact spatial support of the
  mode rather than by the dark-matter density alone. The oscillation is confined to the inner dark core and responds to the total gravitating mass enclosed
  within that region, giving it a shorter dynamical timescale than the
  NM-led mode. Across the plotted fixed-$F_{\rm DM}$ sequences, both branches show an
  overall increase in frequency with mass as the configurations become more
  compact. At a common mass, increasing $F_{\rm DM}$ shifts both branches
  upward: the NM-led shift follows the contraction of $R_{\rm NM}$ in
  \cref{mass_radius}, whereas the DM-led shift also reflects the restructuring
  of the dark core and its coupling to the shared metric, and therefore cannot
  be reduced to the total stellar compactness alone.

\begin{figure*}[!tb]
  \centering
    \includegraphics[width=0.85\linewidth]{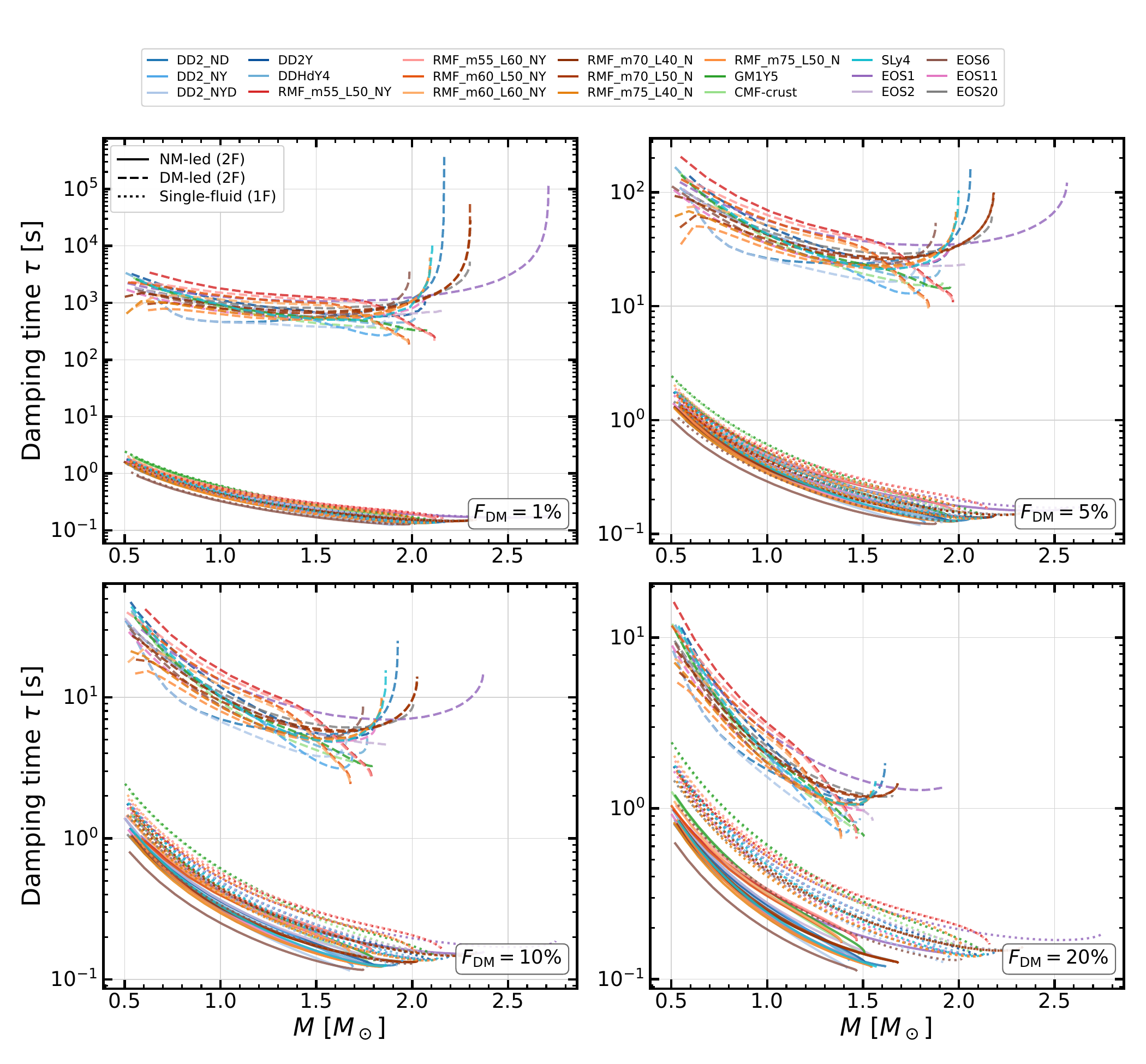}  
 \caption{Full general-relativistic gravitational-wave damping times
  $\tau_{\rm GW}$ of the two-fluid $f$-mode branches in dark-matter-core
  configurations. The four panels correspond to
  $F_{\rm DM}=1\%,5\%,10\%$, and $20\%$, and use logarithmic vertical scales.
  Colors identify the nuclear EOS listed in the legend. Solid and dashed curves
  denote the NM-led and DM-led two-fluid branches, respectively, while dotted
  curves show the corresponding pure-normal-matter single-fluid reference
  sequences.}

  \label{tau_nm}
\end{figure*}

Across the EOS ensemble at fixed $F_{\rm DM}$, both branches broadly follow
  the normal-matter compactness ordering: models with smaller $R_{1.4}$
  generally produce higher frequencies. This ordering reflects the complete
  density and sound-speed profiles rather than the presence of hyperons or
  $\Delta$ isobars alone. At $F_{\rm DM}=10\%$, the nucleonic EOS6 model
  ($R_{1.4}=10.87\,\mathrm{km}$) gives the highest NM-led frequency, while
  DD2\_NYD ($R_{1.4}=11.26\,\mathrm{km}$) gives the marginally highest DM-led
  frequency; the large-radius RMF\_m55\_L60\_NY model
  ($R_{1.4}=13.22\,\mathrm{km}$) gives the lowest frequency on both branches.

  Across the RMF ladder, the mean DM-led frequency of the two $L$ variants
  increases by approximately $23\%$ between $m^*/m_N=0.55$ and $0.75$ at
  $F_{\rm DM}=1\%$--$10\%$. Because the low-$m^*$ models are hyperonic and
  the high-$m^*$ models are nucleonic, this difference represents the combined
  effects of effective mass and composition and cannot be attributed to
  $m^*/m_N$ alone. The controlled comparison is provided by the fixed-$m^*$
  pairs: varying $L$ changes the DM-led frequency by at most $0.18\%$ across
  all pairs that support a $1.4\,M_\odot$ configuration, despite
  $R_{1.4}$ differences as large as $0.77\,\mathrm{km}$. Thus, within this
  ensemble, the DM-led frequency is only weakly sensitive to the
  near-saturation EOS variation encoded by $L$.

  As summarized in \cref{tab:core_config_full_gr}, the $1.4\,M_\odot$
  frequency bands rise from $1.48$--$1.96\,\mathrm{kHz}$ (NM-led) and
  $2.25$--$2.93\,\mathrm{kHz}$ (DM-led) at $F_{\rm DM}=1\%$ to
  $1.98$--$2.80\,\mathrm{kHz}$ and $3.13$--$4.63\,\mathrm{kHz}$,
  respectively, at $20\%$, among the EOS that still support a
  $1.4\,M_\odot$ configuration. The branch labels identify the dominant fluid
  motion rather than dynamically isolated oscillations: both frequencies depend
  on the complete two-fluid structure, so neither the dark-matter fraction nor
  the nuclear EOS can be inferred from a single frequency alone.

Although the two branches remain comparatively close in frequency,
  \cref{tau_nm} shows that their gravitational-wave damping times respond very
  differently to the dark-matter fraction. At $1.4\,M_\odot$, the
  ensemble-median ratio $\tau_{\rm DM}/\tau_{\rm NM}$ decreases from
  approximately $2.6\times10^{3}$ at $F_{\rm DM}=1\%$ to
  $1.2\times10^{2}$, $30$, and $7.5$ at $5\%$, $10\%$, and $20\%$,
  respectively. The damping-time separation is therefore largest in the
  low-fraction regime and narrows as the dark component becomes dynamically
  more important. In the leading mass-quadrupole diagnostic,
  $\tau^{-1}\propto\omega^{4}|Q|^{2}/J$, so the damping measures the net
  coupling of a mode to the radiative channel. At $1.4\,M_\odot$, the NM-led
  damping time remains within
  $\tau_{\rm NM}^{1.4}\simeq0.12$--$0.33\,\mathrm{s}$ and varies much less
  strongly with $F_{\rm DM}$ than the DM-led damping time.

  The DM-led branch is much more sensitive to the dark-matter fraction.
  As summarized in \cref{tab:core_config_full_gr},
  $\tau_{\rm DM}^{1.4}$ decreases from $395$--$1317\,\mathrm{s}$ at
  $F_{\rm DM}=1\%$ to $18$--$44\,\mathrm{s}$, $4$--$9\,\mathrm{s}$, and
  $0.7$--$1.6\,\mathrm{s}$ at $5\%$, $10\%$, and $20\%$, respectively.
  Logarithmic fits give
  $\tau_{\rm DM}^{1.4}\propto F_{\rm DM}^{-n}$ with
  $n=2.03$--$2.34$ across the ensemble; nineteen EOS contribute all four
  fractions, while the two RMF\_m60 hyperonic models contribute the three
  fractions for which they support a $1.4\,M_\odot$ star. The approximate
  $F_{\rm DM}^{-2}$ behavior is therefore an empirical ensemble trend rather
  than an exact consequence of $Q_\alpha\propto F_{\rm DM}$. As the fraction
  increases, the mode frequency, inertia, eigenfunction, and total quadrupole
  all change; the fitted exponent summarizes the resulting fraction dependence
  of the full-GR radiative damping time.

  At $F_{\rm DM}=1\%$, the canonical EOS1 model has
  $\tau_{\rm DM}/\tau_{\rm NM}\simeq3.6\times10^{3}$, demonstrating the weak
  projection of the DM-led mode onto the outgoing gravitational-wave channel.
  The representative profiles in
  \cref{fig:core_eigenfunctions,fig:core_edge} show that the DM-led motion is
  localized within the dark core, induces only a small normal-fluid response,
  and produces metric perturbations that become strongly suppressed before the
  outer surface. Across the EOS ensemble, more compact configurations broadly
  tend to have shorter damping times, but the spread also depends on the
  two-fluid eigenfunction and its surface projection onto the outgoing Zerilli
  solution. Away from the high-mass endpoint upturns discussed below,
  $\tau_{\rm DM}$ generally decreases with mass as the frequency and radiative
  coupling increase.

Near the maximum-mass endpoint, some DM-led sequences develop an
  EOS-dependent upturn in $\tau_{\rm DM}$: the mode radiates gravitational
  waves less efficiently. Its strength is closely ordered by
  the endpoint compactness, with Spearman coefficients
  $r_s=0.85$--$0.98$ across the four dark-matter fractions, and the effect
  becomes less common as $F_{\rm DM}$ increases. For every sequence satisfying
  $\tau_{\rm DM}(M_{\rm end})/\min_M\tau_{\rm DM}>1.5$, the
  leading-quadrupole diagnostic
  $R_Q=|Q_{\rm N}+Q_{\rm D}|/(|Q_{\rm N}|+|Q_{\rm D}|)$ increases toward the
  endpoint. The upturns are therefore not associated with the low-mass
  source-cancellation peak of Ref.~\cite{Zheng:2026fwf}; instead, they are
  consistent with the high-mass surface-radiative-projection minimum identified
  in that work, where the matter quadrupole remains nonzero but its coupling to
  the outgoing Zerilli solution is strongly suppressed. The quadrupole estimate
  accounts for part of the rise in some EOS but substantially underpredicts the
  strongest cases. None of the pure-NM reference sequences satisfies the same
  criterion, although mild endpoint curvature can occur.

%%%%%%%%%%%%%%%%%%%%%%%%%%%%%%%%%%%%%%%%%%%%%%%%%%%%%%%%%%%%%%%%%%%%%%%%%%%%%%%%%%%%%%%%
\subsection{Halo}
\label{sec:halo_results}

\begin{figure}[!htb]
  \centering
    \includegraphics[width=\linewidth]{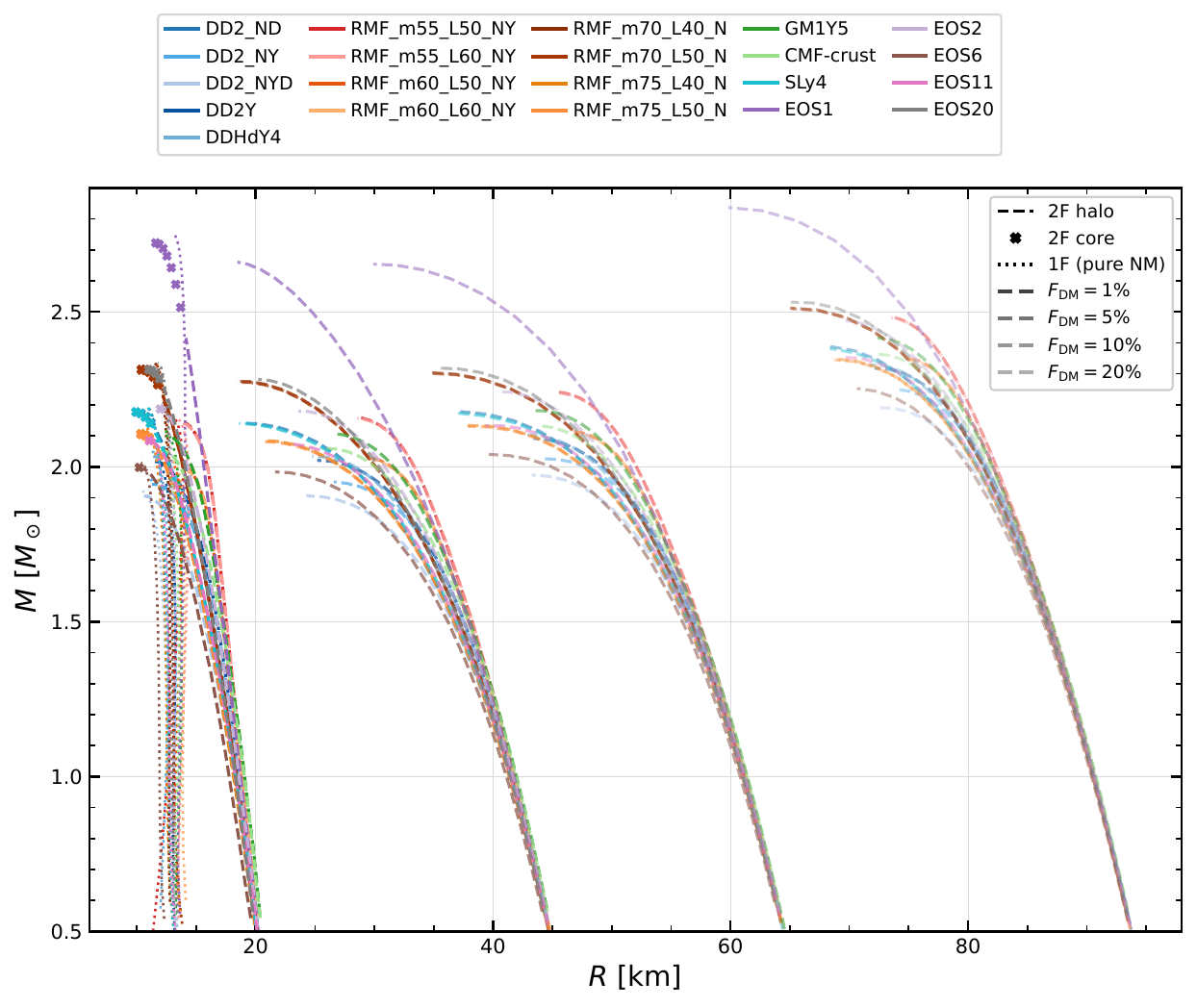}
\caption{Mass--radius relations for the 21 nuclear EOS.
    Dashed curves denote two-fluid halo configurations
    ($R_{\rm DM}>R_{\rm NM}$), dotted curves denote the corresponding
    pure normal-matter single-fluid sequences, and crosses mark
    two-fluid core configurations ($R_{\rm DM}<R_{\rm NM}$).
    The plotted radius is $R_{\rm DM}$ for two-fluid configurations
    and $R_{\rm NM}$ for single-fluid sequences. Colors identify
    the nuclear EOS. For each EOS, the two-fluid curves from darkest
    to lightest correspond to $F_{\rm DM}=1\%,\,5\%,\,10\%$,
    and $20\%$, respectively.}
  \label{mass_halo}
\end{figure}

\cref{mass_halo} compares pure-NM sequences, plotted against $R_{\rm NM}$,
  with two-fluid sequences plotted against $R_{\rm DM}$. At fixed total mass,
  increasing $F_{\rm DM}$ increases the dark mass and reduces the NM mass,
  while repulsive bosonic pressure supports an increasingly extended halo.
  In the dilute Newtonian limit, treating the NM component as a point mass
  and neglecting halo self-gravity, the low-density relation
  $p_\chi\propto\epsilon_\chi^2$ gives
  $R_{\rm DM}\propto\sqrt{F_{\rm DM}/(1-F_{\rm DM})}$.
  This limiting estimate suggests roughly fivefold growth from
  $F_{\rm DM}=1\%$ to $20\%$, consistent with the increase from
  $\approx17$ to $\approx88$~km at $1.4\,M_\odot$ in
  \cref{tab:dm_fmode_summary}. The full-GR equilibrium solutions retain
  both the finite NM radius and halo self-gravity. The nuclear-EOS spread
  at a given mass narrows as the halo grows: beyond the NM surface, the
  nuclear component contributes through its enclosed mass, while the dark
  distribution responds to its own EOS and the combined gravitational field.
  The halo radius thus becomes less sensitive to the detailed nuclear
  density profile. The maximum mass varies nonmonotonically at small dark fractions:
  11 of the 21 EOS decrease slightly from $F_{\rm DM}=1\%$ to $5\%$,
  whereas 20 increase from $5\%$ to $10\%$ and all increase from $10\%$
  to $20\%$. For every EOS, $M_{\max}(20\%)>M_{\max}(1\%)$.
  This contrasts with the core trend in \cref{mass_radius} and agrees
  qualitatively with previous bosonic-DM studies~\cite{Karkevandi:2024vov}.
  At the respective sequence maxima, the increasing mass of the extended,
  pressure-supported dark component outweighs the decreasing NM contribution;
  the NM component need not remain at its pure-NM maximum mass. The $1\%$ family lies close to the core--halo boundary. Some sequences
  cross to $R_{\rm DM}<R_{\rm NM}$ as the stellar mass increases, with
  crosses marking these core configurations. The transition occurs at an
  EOS-dependent mass, bracketed by the last halo point and the first core
  point. By contrast, $R_{\rm DM}>R_{\rm NM}$ throughout the displayed
  $F_{\rm DM}=5$--$20\%$ sequences.

\begin{figure*}[!htb]
  \centering
    \includegraphics[width=0.85\linewidth]{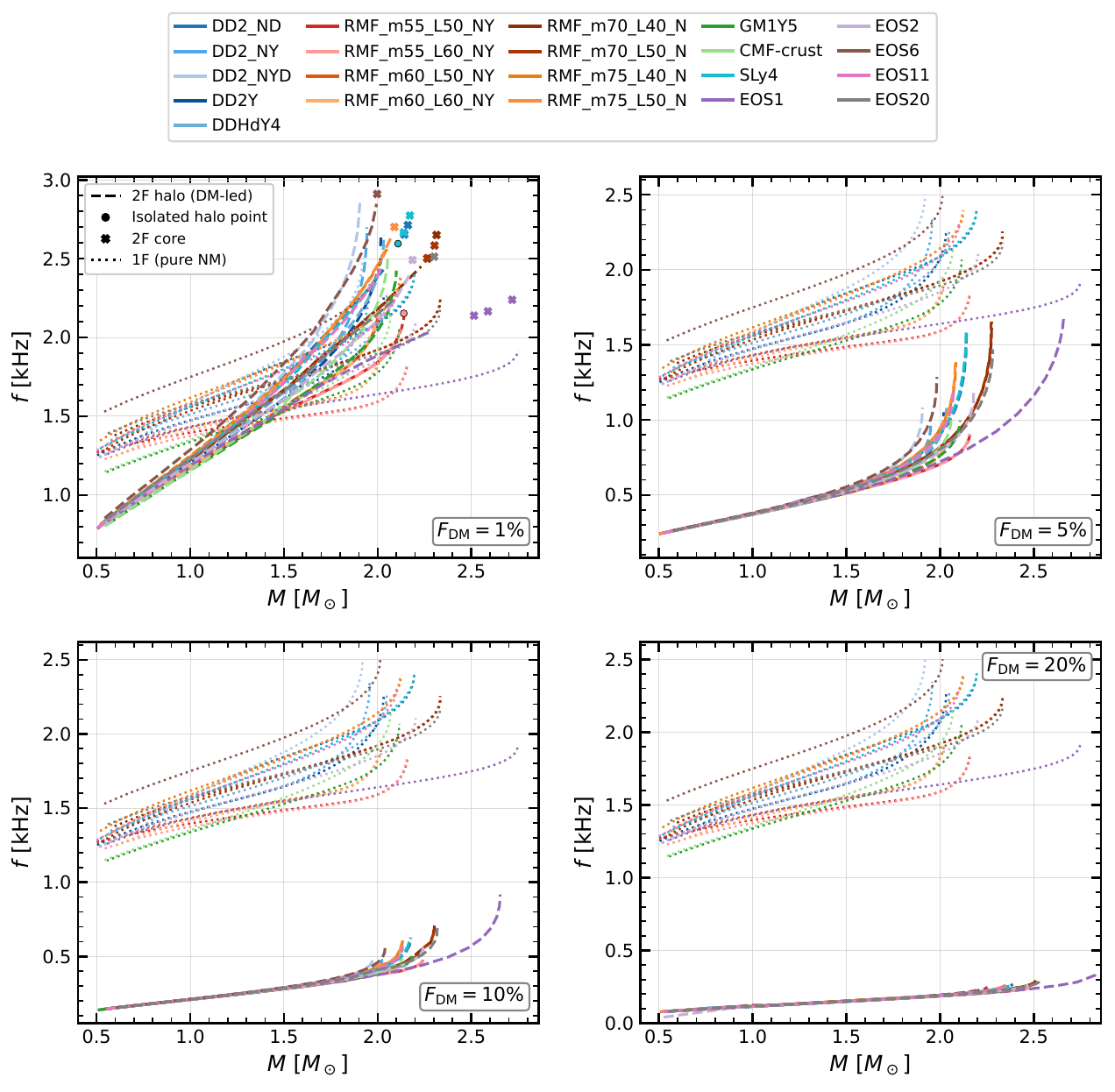}
\caption{DM-led quadrupolar $f$-mode frequencies versus gravitational
    mass for the 21 nuclear EOS. The panels show
    $F_{\rm DM}=1\%,\,5\%,\,10\%$, and $20\%$ in reading order.
    Dashed curves denote two-fluid halo configurations
    ($R_{\rm DM}>R_{\rm NM}$), and crosses mark two-fluid core
    configurations ($R_{\rm DM}<R_{\rm NM}$). Filled circles mark
    resolved halo points that are not joined across unresolved gaps.
    Dotted curves show the corresponding pure normal-matter
    single-fluid $f$-mode frequencies. Colors identify the nuclear EOS.}
  \label{halo_freq}
\end{figure*}

 \cref{halo_freq} shows the full-GR quadrupolar dark-matter-led $f$-mode
  frequencies along the halo-study sequences of \cref{mass_halo}. The NM-led
  two-fluid branch is not shown; the dotted curves are separate pure-NM
  single-fluid references. The DM-led mode is supported predominantly by the
  extended dark fluid, while its restoring force combines the dark pressure
  response with the gravitational field generated by both components. For
  extended halos, its characteristic frequency approximately follows the
  dynamical scale $f_{\rm DM}\propto\sqrt{M/R_{\rm DM}^{3}}$. Along
  fixed-$F_{\rm DM}$ sequences, the overall frequency increase with stellar
  mass accompanies halo contraction and a stronger gravitational field,
  producing the upward curvature toward the endpoints. At fixed total mass,
  increasing $F_{\rm DM}$ instead expands the halo and lengthens its dynamical timescale, lowering the frequency despite the larger dark-matter content.

At $1.4\,M_\odot$, \cref{tab:dm_fmode_summary} shows that the nearly fivefold increase in
$R_{\rm DM}$ between $F_{\rm DM}=1\%$ and $20\%$ is accompanied by an
approximately tenfold decrease in frequency, from $1.455$--$1.713$~kHz to
$0.142$--$0.146$~kHz, consistent with the approximate $R_{\rm DM}^{-3/2}$
dependence. The nuclear EoS spread narrows with increasing $F_{\rm DM}$
because a well-developed halo is governed mainly by the common
dark-matter EoS, whereas the normal-matter EoS affects its oscillation
only indirectly through the compact inner component and the shared
metric; the larger spread near the endpoints primarily reflects the
different EoS-dependent maximum masses and corresponding degrees of halo
compression. The $1\%$ family remains exceptional because it lies close
to the core--halo boundary, where the dark and normal radii are
comparable and the mode retains greater sensitivity to the internal
two-fluid structure. The cross-marked core configurations should therefore not
be read off the extended-halo scaling. Intersections with the
pure-normal-matter reference curves likewise do not represent mode
crossings, since those curves belong to different single-fluid
equilibrium backgrounds. This interpretation is consistent with previous
two-fluid studies of low-frequency dark-matter-led halo
oscillations~\cite{Sotani:2025lzy}. 

The same structural evolution also controls the radiative damping of this
  branch. Using the extended-precision outgoing-wave calculation, we resolve
  DM-led complex QNMs for all $84$ configurations at $1.4\,M_\odot$ and all
  $73$ maximum-mass configurations that remain halos. At $1.4\,M_\odot$,
  $\tau_{\rm DM}^{1.4}$ increases from $0.46$--$8.03$~s at
  $F_{\rm DM}=1\%$ to approximately $450$~s at $20\%$, with the intermediate
  fractions lying between these ranges
  (\cref{tab:dm_fmode_summary}). The expanding halo's lower frequency is
  accompanied by weaker radiative coupling, opposite to the core trend.
  The frequency, mode inertia, and radiative projection change together, so
  a frequency-power scaling alone does not determine the damping time.
  The narrowing EOS spread parallels the frequency convergence and reflects
  the reduced sensitivity to nuclear structure in the extended-halo regime.
  Toward maximum mass, halo contraction raises the frequency and strengthens
  the radiative coupling for $F_{\rm DM}=5$--$20\%$, shortening the damping
  time by factors of several to a few tens relative to its
  $1.4\,M_\odot$ value.

  The $1\%$ endpoints behave differently. Ten remain halos, and all have
  $R_Q=|Q_{\rm N}+Q_{\rm D}|/(|Q_{\rm N}|+|Q_{\rm D}|)\leq0.096$,
  indicating strong cancellation between the component matter quadrupoles.
  Their full-GR damping times range from approximately $700$~s to
  $1.20\times10^{6}$~s, with the longest lifetime occurring for
  RMF\_m55\_L50\_NY. Here $R_Q$ diagnoses source cancellation, while the
  lifetime is determined by the complete outgoing-wave solution.
  The other eleven maximum-mass endpoints are core configurations and are
  therefore not assigned halo damping times in
  \cref{tab:dm_fmode_summary}. These are isolated-star, ideal-fluid radiative
  damping times; observability additionally depends on the mode-excitation
  amplitude and detector response.

\clearpage
\makeatletter
\@ifundefined{onecolumngrid}{}{\onecolumngrid}
\makeatother
\begingroup
\setlength{\LTleft}{0pt}
\setlength{\LTright}{0pt}
\setlength{\LTcapwidth}{\textwidth}
\setlength{\tabcolsep}{1.35pt}
\renewcommand{\arraystretch}{0.82}
\tiny
\begin{longtable}{@{}l c r r r r r r r r r r r@{}}
\caption{Core-configuration stellar properties and full general-relativistic
  two-fluid $f$-mode results for dark-matter-admixed neutron stars.
  $F_{\rm DM}$ is the dark-matter fraction; $M_{\max}$ and $R_{\max}$
  are evaluated at the maximum-mass configuration, while $R_{1.4}$ is
  the radius at $1.4\,M_\odot$. The frequencies are given in kHz and
  the damping times in seconds. NM and DM denote the normal-matter-led
  and dark-matter-led branches, respectively. Rows labelled ``NM'' in
  the $F_{\rm DM}$ column give the pure normal-matter single-fluid
  reference configurations ($F_{\rm DM}=0$).}
  \label{tab:core_config_full_gr}\\

\toprule
\multicolumn{2}{c}{Model} & \multicolumn{3}{c}{Structure} & \multicolumn{4}{c}{$f$-mode frequency} & \multicolumn{4}{c}{Damping time} \\
\cmidrule(lr){1-2}\cmidrule(lr){3-5}\cmidrule(lr){6-9}\cmidrule(l){10-13}
EoS & $F_{\rm DM}$ & $M_{\max}$ & $R_{\max}$ & $R_{1.4}$ & $f_{\rm NM}^{1.4}$ & $f_{\rm DM}^{1.4}$ & $f_{\rm NM}^{\max}$ & $f_{\rm DM}^{\max}$ & $\tau_{\rm NM}^{1.4}$ & $\tau_{\rm DM}^{1.4}$ & $\tau_{\rm NM}^{\max}$ & $\tau_{\rm DM}^{\max}$ \\
 & (\%) & ($M_\odot$) & (km) & (km) & (kHz) & (kHz) & (kHz) & (kHz) & (s) & (s) & (s) & (s) \\
\midrule
\endfirsthead
\toprule
\multicolumn{2}{c}{Model} & \multicolumn{3}{c}{Structure} & \multicolumn{4}{c}{$f$-mode frequency} & \multicolumn{4}{c}{Damping time} \\
\cmidrule(lr){1-2}\cmidrule(lr){3-5}\cmidrule(lr){6-9}\cmidrule(l){10-13}
EoS & $F_{\rm DM}$ & $M_{\max}$ & $R_{\max}$ & $R_{1.4}$ & $f_{\rm NM}^{1.4}$ & $f_{\rm DM}^{1.4}$ & $f_{\rm NM}^{\max}$ & $f_{\rm DM}^{\max}$ & $\tau_{\rm NM}^{1.4}$ & $\tau_{\rm DM}^{1.4}$ & $\tau_{\rm NM}^{\max}$ & $\tau_{\rm DM}^{\max}$ \\
 & (\%) & ($M_\odot$) & (km) & (km) & (kHz) & (kHz) & (kHz) & (kHz) & (s) & (s) & (s) & (s) \\
\midrule
\endhead
\midrule
\multicolumn{13}{r}{\textit{continued on next page}}\\
\endfoot
\bottomrule
\endlastfoot
EOS1 & NM & 2.748 & 13.1 & 13.77 & 1.526 & -- & 1.92 & -- & 0.3085 & -- & 0.1872 & -- \\
 & 1 & 2.712 & 13 & 13.68 & 1.547 & 2.342 & 1.942 & 2.331 & 0.3024 & 1082 & 0.1836 & $7845^{\dagger}$ \\
 & 5 & 2.564 & 12.69 & 13.35 & 1.629 & 2.522 & 2.014 & 2.609 & 0.2714 & 39.66 & 0.1697 & 121.7 \\
 & 10 & 2.371 & 12.29 & 12.93 & 1.737 & 2.712 & 2.109 & 3.007 & 0.2398 & 8.466 & 0.1561 & 14.93 \\
 & 20 & 1.92 & 11.71 & 12.07 & 1.977 & 3.132 & 2.226 & 4.037 & 0.1937 & 1.59 & 0.1424 & 1.295 \\
\addlinespace[1pt]
EOS2 & NM & 2.204 & 12.03 & 13.22 & 1.614 & -- & 2.117 & -- & 0.2746 & -- & 0.1441 & -- \\
 & 1 & 2.171 & 11.95 & 13.12 & 1.639 & 2.526 & 2.137 & 2.931 & 0.2668 & 689.9 & 0.1422 & 724.1 \\
 & 5 & 2.037 & 11.67 & 12.75 & 1.74 & 2.732 & 2.21 & 3.222 & 0.2369 & 27.94 & 0.1368 & 23.31 \\
 & 10 & 1.872 & 11.32 & 12.27 & 1.874 & 2.964 & 2.296 & 3.577 & 0.2072 & 6.094 & 0.1331 & 4.581 \\
 & 20 & 1.551 & 10.66 & 11.23 & 2.199 & 3.564 & 2.445 & 4.27 & 0.1632 & 1.117 & 0.136 & 0.8473 \\
\addlinespace[1pt]
EOS6 & NM & 2.014 & 10.35 & 11.82 & 1.926 & -- & 2.503 & -- & 0.1938 & -- & 0.132 & -- \\
 & 1 & 1.986 & 10.25 & 11.72 & 1.957 & 2.925 & 2.538 & 3.195 & 0.1884 & 574.4 & 0.1301 & 3550 \\
 & 5 & 1.879 & 9.986 & 11.35 & 2.083 & 3.151 & 2.638 & 3.506 & 0.1683 & 24.01 & 0.1233 & 53.9 \\
 & 10 & 1.744 & 9.697 & 10.87 & 2.252 & 3.416 & 2.751 & 3.9 & 0.1485 & 5.327 & 0.1169 & 8.604 \\
 & 20 & 1.465 & 9.286 & 9.73 & 2.697 & 4.207 & 2.901 & 4.716 & 0.12 & 1.064 & 0.1128 & 1.165 \\
\addlinespace[1pt]
EOS11 & NM & 2.101 & 11.1 & 12.66 & 1.745 & -- & 2.33 & -- & 0.2348 & -- & 0.1367 & -- \\
 & 1 & 2.071 & 11.03 & 12.55 & 1.774 & 2.725 & 2.353 & 3.084 & 0.2278 & 541.2 & 0.1348 & 1348 \\
 & 5 & 1.953 & 10.76 & 12.16 & 1.89 & 2.943 & 2.437 & 3.388 & 0.2022 & 23.39 & 0.1285 & 33.88 \\
 & 10 & 1.805 & 10.54 & 11.66 & 2.043 & 3.195 & 2.509 & 3.729 & 0.1768 & 5.285 & 0.1238 & 5.964 \\
 & 20 & 1.507 & 9.984 & 10.54 & 2.428 & 3.882 & 2.667 & 4.494 & 0.14 & 1.028 & 0.1237 & 0.9709 \\
\addlinespace[1pt]
EOS20 & NM & 2.334 & 11.77 & 12.96 & 1.704 & -- & 2.175 & -- & 0.2463 & -- & 0.1532 & -- \\
 & 1 & 2.301 & 11.65 & 12.87 & 1.729 & 2.603 & 2.207 & 2.768 & 0.2399 & 821.5 & 0.1509 & 5296 \\
 & 5 & 2.172 & 11.37 & 12.52 & 1.828 & 2.799 & 2.288 & 3.063 & 0.2145 & 31.38 & 0.1415 & 61.33 \\
 & 10 & 2.006 & 11.07 & 12.07 & 1.958 & 3.015 & 2.379 & 3.443 & 0.1902 & 6.838 & 0.1329 & 9.045 \\
 & 20 & 1.658 & 10.52 & 11.12 & 2.259 & 3.534 & 2.528 & 4.299 & 0.1539 & 1.3 & 0.1271 & 1.177 \\
\addlinespace[1pt]
DD2Y & NM & 2.099 & 11.42 & 12.81 & 1.617 & -- & 2.229 & -- & 0.273 & -- & 0.1372 & -- \\
 & 1 & 2.069 & 11.33 & 12.63 & 1.642 & 2.536 & 2.251 & 3.079 & 0.2654 & 758.4 & 0.1354 & 1060 \\
 & 5 & 1.95 & 10.96 & 12.27 & 1.75 & 2.768 & 2.356 & 3.403 & 0.2339 & 28.84 & 0.1293 & 35.24 \\
 & 10 & 1.802 & 10.62 & 11.79 & 1.901 & 3.044 & 2.458 & 3.787 & 0.2015 & 6.207 & 0.1244 & 6.635 \\
 & 20 & 1.507 & 10.06 & 10.66 & 2.311 & 3.82 & 2.611 & 4.523 & 0.1493 & 1.123 & 0.1262 & 1.059 \\
\addlinespace[1pt]
DDHdY4 & NM & 2.049 & 11.2 & 12.66 & 1.665 & -- & 2.286 & -- & 0.2576 & -- & 0.1344 & -- \\
 & 1 & 2.019 & 11.16 & 12.56 & 1.692 & 2.617 & 2.298 & 3.152 & 0.2493 & 603.6 & 0.1331 & 602.2 \\
 & 5 & 1.905 & 10.82 & 12.19 & 1.804 & 2.84 & 2.396 & 3.459 & 0.2204 & 24.44 & 0.1283 & 22.99 \\
 & 10 & 1.765 & 10.33 & 11.7 & 1.958 & 3.111 & 2.546 & 3.88 & 0.1908 & 5.198 & 0.1241 & 5.126 \\
 & 20 & 1.484 & 9.789 & 10.48 & 2.382 & 3.941 & 2.686 & 4.716 & 0.1456 & 0.8954 & 0.1263 & 0.8756 \\
\addlinespace[1pt]
GM1Y5 & NM & 2.116 & 12.32 & 13.78 & 1.49 & -- & 2.061 & -- & 0.3238 & -- & 0.1448 & -- \\
 & 1 & 2.083 & 12.14 & 13.67 & 1.515 & 2.412 & 2.106 & 3.051 & 0.3126 & 584 & 0.1424 & 340 \\
 & 5 & 1.956 & 11.79 & 13.26 & 1.616 & 2.622 & 2.186 & 3.353 & 0.2754 & 25.42 & 0.1388 & 14.91 \\
 & 10 & 1.803 & 11.39 & 12.75 & 1.75 & 2.862 & 2.28 & 3.721 & 0.2373 & 5.621 & 0.1372 & 3.352 \\
 & 20 & 1.507 & 10.65 & 11.49 & 2.119 & 3.629 & 2.442 & 4.522 & 0.1778 & 0.8986 & 0.1437 & 0.6891 \\
\addlinespace[1pt]
CMF-crust & NM & 2.069 & 11.91 & 13.6 & 1.522 & -- & 2.149 & -- & 0.3095 & -- & 0.1398 & -- \\
 & 1 & 2.038 & 11.74 & 13.44 & 1.55 & 2.49 & 2.187 & 3.119 & 0.2985 & 485.7 & 0.1379 & 376.2 \\
 & 5 & 1.916 & 11.43 & 13 & 1.663 & 2.721 & 2.266 & 3.419 & 0.2601 & 21.5 & 0.1344 & 16.16 \\
 & 10 & 1.767 & 11.07 & 12.44 & 1.815 & 2.993 & 2.355 & 3.782 & 0.2219 & 4.76 & 0.133 & 3.528 \\
 & 20 & 1.478 & 10.45 & 11.1 & 2.226 & 3.804 & 2.487 & 4.488 & 0.1649 & 0.8393 & 0.1407 & 0.7096 \\
\addlinespace[1pt]
SLy4 & NM & 2.134 & 10.67 & 12.49 & 1.703 & -- & 2.417 & -- & 0.2458 & -- & 0.1425 & -- \\
 & 1 & 2.106 & 10.6 & 12.37 & 1.733 & 2.712 & 2.442 & 2.99 & 0.2377 & 553.8 & 0.1403 & $8602^{\dagger}$ \\
 & 5 & 1.999 & 10.27 & 11.97 & 1.858 & 2.946 & 2.554 & 3.289 & 0.2082 & 23.97 & 0.1335 & 100.5 \\
 & 10 & 1.862 & 9.888 & 11.46 & 2.024 & 3.215 & 2.689 & 3.697 & 0.1789 & 5.347 & 0.1263 & 15.55 \\
 & 20 & 1.566 & 9.605 & 10.34 & 2.429 & 3.907 & 2.778 & 4.657 & 0.1386 & 1.059 & 0.1189 & 1.445 \\
\addlinespace[1pt]
DD2\_ND & NM & 2.195 & 10.74 & 12.46 & 1.8 & -- & 2.407 & -- & 0.2214 & -- & 0.1503 & -- \\
 & 1 & 2.168 & 10.73 & 12.36 & 1.828 & 2.783 & 2.418 & 2.887 & 0.2153 & 519.4 & 0.1466 & $9999^{\dagger}$ \\
 & 5 & 2.06 & 10.39 & 11.99 & 1.939 & 2.986 & 2.531 & 3.17 & 0.193 & 23.88 & 0.1392 & 161 \\
 & 10 & 1.925 & 9.945 & 11.51 & 2.087 & 3.216 & 2.688 & 3.55 & 0.1707 & 5.572 & 0.132 & 25.18 \\
 & 20 & 1.615 & 9.664 & 10.5 & 2.439 & 3.818 & 2.802 & 4.733 & 0.1384 & 1.137 & 0.1193 & 1.842 \\
\addlinespace[1pt]
DD2\_NY & NM & 1.958 & 11.21 & 13.27 & 1.61 & -- & 2.325 & -- & 0.2757 & -- & 0.1311 & -- \\
 & 1 & 1.93 & 11.11 & 13.17 & 1.637 & 2.549 & 2.352 & 3.295 & 0.2667 & 603.1 & 0.1298 & 405.3 \\
 & 5 & 1.822 & 10.79 & 12.74 & 1.753 & 2.807 & 2.447 & 3.605 & 0.2334 & 22.2 & 0.1262 & 17.75 \\
 & 10 & 1.689 & 10.45 & 12.13 & 1.926 & 3.147 & 2.551 & 3.982 & 0.1976 & 4.345 & 0.1244 & 3.931 \\
 & 20 & 1.426 & 9.934 & 10.39 & 2.501 & 4.317 & 2.684 & 4.747 & 0.1401 & 0.7258 & 0.1302 & 0.753 \\
\addlinespace[1pt]
DD2\_NYD & NM & 1.921 & 10.38 & 12.4 & 1.816 & -- & 2.534 & -- & 0.2174 & -- & 0.126 & -- \\
 & 1 & 1.895 & 10.35 & 12.29 & 1.849 & 2.864 & 2.547 & 3.357 & 0.2103 & 395.3 & 0.1244 & 986.1 \\
 & 5 & 1.794 & 10.03 & 11.85 & 1.982 & 3.111 & 2.664 & 3.672 & 0.1853 & 18.31 & 0.1198 & 32.47 \\
 & 10 & 1.67 & 9.683 & 11.26 & 2.17 & 3.423 & 2.792 & 4.065 & 0.1601 & 4.129 & 0.116 & 6.604 \\
 & 20 & 1.415 & 9.324 & 9.587 & 2.798 & 4.628 & 2.913 & 4.882 & 0.1206 & 0.8792 & 0.1171 & 0.9857 \\
\addlinespace[3pt]
\midrule
RMF\_m55\_L50\_NY & NM & 2.159 & 13.44 & 14.17 & 1.47 & -- & 1.848 & -- & 0.3339 & -- & 0.16 & -- \\
 & 1 & 2.119 & 12.98 & 13.21 & 1.49 & 2.253 & 1.85 & 2.856 & 0.3241 & 1317 & 0.1604 & 230.4 \\
 & 5 & 1.969 & 12.57 & 12.92 & 1.569 & 2.433 & 1.92 & 3.149 & 0.2909 & 43.64 & 0.1579 & 10.84 \\
 & 10 & 1.789 & 12.12 & 12.56 & 1.674 & 2.633 & 1.99 & 3.444 & 0.2565 & 8.735 & 0.1595 & 2.609 \\
 & 20 & 1.467 & 11.18 & 11.52 & 1.995 & 3.406 & 2.145 & 3.948 & 0.1937 & 1.021 & 0.1697 & 0.6701 \\
\addlinespace[1pt]
RMF\_m55\_L60\_NY & NM & 2.156 & 13.8 & 15.08 & 1.459 & -- & 1.847 & -- & 0.3391 & -- & 0.1603 & -- \\
 & 1 & 2.116 & 13.23 & 13.98 & 1.479 & 2.252 & 1.871 & 2.902 & 0.3287 & 1181 & 0.1586 & 214.3 \\
 & 5 & 1.967 & 12.93 & 13.65 & 1.559 & 2.431 & 1.918 & 3.153 & 0.295 & 41.61 & 0.1583 & 10.62 \\
 & 10 & 1.787 & 12.57 & 13.22 & 1.665 & 2.633 & 1.971 & 3.402 & 0.2599 & 8.436 & 0.1618 & 2.68 \\
 & 20 & 1.466 & 11.63 & 12.04 & 1.988 & 3.412 & 2.139 & 3.952 & 0.1957 & 1.001 & 0.1712 & 0.6624 \\
\addlinespace[1pt]
RMF\_m60\_L50\_NY & NM & 2.021 & 13.23 & 14.52 & 1.53 & -- & 1.928 & -- & 0.3065 & -- & 0.1543 & -- \\
 & 1 & 1.984 & 12.67 & 13.5 & 1.552 & 2.357 & 1.951 & 3.051 & 0.2989 & 1035 & 0.153 & 184.4 \\
 & 5 & 1.844 & 12.36 & 13.17 & 1.638 & 2.544 & 2.007 & 3.301 & 0.2669 & 37.38 & 0.1526 & 9.719 \\
 & 10 & 1.678 & 11.94 & 12.72 & 1.76 & 2.791 & 2.085 & 3.594 & 0.2331 & 6.996 & 0.1542 & 2.431 \\
 & 20 & 1.387 & 11.14 & -- & -- & -- & 2.234 & 4.022 & -- & -- & 0.1655 & 0.6697 \\
\addlinespace[1pt]
RMF\_m60\_L60\_NY & NM & 2.016 & 13.41 & 14.98 & 1.519 & -- & 1.927 & -- & 0.3113 & -- & 0.1547 & -- \\
 & 1 & 1.979 & 12.83 & 13.89 & 1.54 & 2.354 & 1.951 & 3.056 & 0.3039 & 949 & 0.1533 & 180.7 \\
 & 5 & 1.84 & 12.54 & 13.53 & 1.627 & 2.543 & 2.006 & 3.307 & 0.2707 & 35.54 & 0.1531 & 9.545 \\
 & 10 & 1.675 & 12.12 & 13.04 & 1.75 & 2.793 & 2.083 & 3.601 & 0.236 & 6.724 & 0.155 & 2.392 \\
 & 20 & 1.385 & 11.36 & -- & -- & -- & 2.227 & 4.028 & -- & -- & 0.1672 & 0.6596 \\
\addlinespace[1pt]
RMF\_m70\_L40\_N & NM & 2.336 & 11.62 & 13.59 & 1.7 & -- & 2.245 & -- & 0.2471 & -- & 0.1579 & -- \\
 & 1 & 2.305 & 11.4 & 12.69 & 1.726 & 2.621 & 2.248 & 2.732 & 0.2397 & 730.3 & 0.1537 & $9072^{\dagger}$\\
 & 5 & 2.183 & 11.11 & 12.34 & 1.829 & 2.823 & 2.338 & 3.023 & 0.2151 & 28.73 & 0.1441 & 99.87 \\
 & 10 & 2.027 & 10.72 & 11.89 & 1.964 & 3.045 & 2.459 & 3.425 & 0.1892 & 6.355 & 0.1349 & 14.06 \\
 & 20 & 1.681 & 10.27 & 10.94 & 2.277 & 3.581 & 2.589 & 4.418 & 0.1526 & 1.241 & 0.1254 & 1.404 \\
\addlinespace[1pt]
RMF\_m70\_L50\_N & NM & 2.333 & 11.69 & 13.89 & 1.69 & -- & 2.246 & -- & 0.2502 & -- & 0.1577 & -- \\
 & 1 & 2.302 & 11.35 & 12.95 & 1.715 & 2.621 & 2.277 & 2.743 & 0.2426 & 661.3 & 0.1554 & $8490^{\dagger}$ \\
 & 5 & 2.18 & 11.18 & 12.57 & 1.82 & 2.824 & 2.339 & 3.026 & 0.2175 & 27.38 & 0.144 & 98.64 \\
 & 10 & 2.024 & 10.79 & 12.1 & 1.956 & 3.047 & 2.459 & 3.428 & 0.191 & 6.136 & 0.1349 & 13.95 \\
 & 20 & 1.68 & 10.37 & 11.1 & 2.27 & 3.584 & 2.587 & 4.418 & 0.1539 & 1.219 & 0.1258 & 1.401 \\
\addlinespace[1pt]
RMF\_m75\_L40\_N & NM & 2.125 & 11.03 & 13.19 & 1.795 & -- & 2.4 & -- & 0.2222 & -- & 0.1403 & -- \\
 & 1 & 2.096 & 10.78 & 12.34 & 1.824 & 2.778 & 2.408 & 3.014 & 0.2157 & 556.3 & 0.1375 & 4754 \\
 & 5 & 1.985 & 10.39 & 11.95 & 1.941 & 2.997 & 2.537 & 3.332 & 0.1917 & 23.64 & 0.1311 & 69.61 \\
 & 10 & 1.843 & 10.19 & 11.46 & 2.098 & 3.248 & 2.614 & 3.702 & 0.1684 & 5.348 & 0.1235 & 9.688 \\
 & 20 & 1.544 & 9.675 & 10.37 & 2.481 & 3.914 & 2.786 & 4.671 & 0.1349 & 1.072 & 0.1185 & 1.268 \\
\addlinespace[1pt]
RMF\_m75\_L50\_N & NM & 2.121 & 11.09 & 13.44 & 1.783 & -- & 2.401 & -- & 0.2252 & -- & 0.1401 & -- \\
 & 1 & 2.093 & 10.72 & 12.55 & 1.813 & 2.779 & 2.442 & 3.031 & 0.2185 & 505.1 & 0.1385 & 6277 \\
 & 5 & 1.982 & 10.55 & 12.14 & 1.931 & 2.998 & 2.507 & 3.32 & 0.194 & 22.51 & 0.1302 & 61.21 \\
 & 10 & 1.84 & 10.16 & 11.62 & 2.089 & 3.251 & 2.643 & 3.733 & 0.17 & 5.163 & 0.1238 & 10.29 \\
 & 20 & 1.542 & 9.752 & 10.49 & 2.476 & 3.921 & 2.784 & 4.673 & 0.1358 & 1.053 & 0.119 & 1.26 \\
\end{longtable}

\par\noindent{\scriptsize
$^{\dagger}$ For the weakly radiating $F_{\rm DM}=1\%$ DM-led branches whose
endpoint damping time exceeds $10^{4}\,{\rm s}$, the marked entry is evaluated
at the highest-mass sampled configuration satisfying
$\tau_{\rm DM}\leq10^{4}\,{\rm s}$. All unmarked quantities carrying a
superscript ``max,'' including $f_{\rm DM}^{\max}$, are evaluated at the
maximum-mass sampled configuration.\par}
\endgroup
\twocolumngrid

\twocolumngrid

\begin{table*}[htbp]
\caption{Maximum masses, dark-matter radii, and full general-relativistic dark-matter-led $f$-mode frequencies and gravitational-wave damping times along the halo-study sequences. $R_{\rm DM}^{\max}$ is evaluated at $M_{\max}$ and $R_{\rm DM}^{1.4}$ at $1.4\,M_\odot$; frequencies are in kHz and damping times in seconds. For the $F_{\rm DM}=1\%$ sequences whose maximum-mass points are marked as core configurations in \cref{mass_halo}, $R_{\rm DM}^{\max}$ is the inner dark-component radius. A dagger marks $f_{\rm DM}^{\max}$ evaluated at the highest-mass configuration with a reliably resolved dark-matter-led mode rather than exactly at $M_{\max}$. The quantity $\tau_{\rm DM}^{\max}$ is reported only when the maximum-mass model remains a halo; a dash denotes a nonhalo maximum-mass endpoint.}
\label{tab:dm_fmode_summary}
\centering
\tiny
\setlength{\tabcolsep}{1.1pt}
\renewcommand{\arraystretch}{0.95}
\resizebox{\textwidth}{!}{%
\begin{tabular}{@{}lcccccccc@{\hspace{0.7em}}lcccccccc@{}}
\toprule
\multicolumn{2}{c}{Model} & \multicolumn{3}{c}{Structure} & \multicolumn{2}{c}{Frequency} & \multicolumn{2}{c}{Damping time} & \multicolumn{2}{c}{Model} & \multicolumn{3}{c}{Structure} & \multicolumn{2}{c}{Frequency} & \multicolumn{2}{c}{Damping time} \\
\cmidrule(lr){1-2}\cmidrule(lr){3-5}\cmidrule(lr){6-7}\cmidrule(lr){8-9}\cmidrule(lr){10-11}\cmidrule(lr){12-14}\cmidrule(lr){15-16}\cmidrule(l){17-18}
EoS & $F_{\rm DM}$ & $M_{\max}$ & $R_{\rm DM}^{\max}$ & $R_{\rm DM}^{1.4}$ & $f_{\rm DM}^{1.4}$ & $f_{\rm DM}^{\max}$ & $\tau_{\rm DM}^{1.4}$ & $\tau_{\rm DM}^{\max}$ & EoS & $F_{\rm DM}$ & $M_{\max}$ & $R_{\rm DM}^{\max}$ & $R_{\rm DM}^{1.4}$ & $f_{\rm DM}^{1.4}$ & $f_{\rm DM}^{\max}$ & $\tau_{\rm DM}^{1.4}$ & $\tau_{\rm DM}^{\max}$ \\
& (\%) & ($M_\odot$) & (km) & (km) & (kHz) & (kHz) & (s) & (s) & & (\%) & ($M_\odot$) & (km) & (km) & (kHz) & (kHz) & (s) & (s) \\
\midrule
\addlinespace[2pt]\multirow{4}{*}{EOS1} & 1 & 2.723 & 11.58 & 17.56 & 1.4730 & $2.2390^{\dagger}$ & 1.843 & -- & \multirow{4}{*}{DD2\_NY} & 1 & 1.943 & 11.27 & 17.16 & 1.5270 & 2.6980 & 3.445 & 1271 \\
 & 5 & 2.661 & 18.45 & 38.78 & 0.4858 & 1.6880 & 91.04 & 3.748 &  & 5 & 1.951 & 26.60 & 38.51 & 0.4903 & 0.9618 & 89.83 & 19.93 \\
 & 10 & 2.654 & 29.93 & 58.05 & 0.2671 & 0.9175 & 193.8 & 11.06 &  & 10 & 2.026 & 44.20 & 57.89 & 0.2681 & 0.4731 & 192.6 & 42.31 \\
 & 20 & 2.837 & 59.82 & 87.74 & 0.1425 & 0.3524 & 446.7 & 29.35 &  & 20 & 2.250 & 74.09 & 87.63 & 0.1428 & 0.2309 & 445.7 & 82.65 \\
\addlinespace[2pt]\multirow{4}{*}{EOS2} & 1 & 2.186 & 11.97 & 17.14 & 1.5290 & 2.4920 & 3.274 & -- & \multirow{4}{*}{DD2\_NYD} & 1 & 1.906 & 10.60 & 16.25 & 1.6530 & 2.8690 & 6.701 & 2519 \\
 & 5 & 2.179 & 23.61 & 38.49 & 0.4909 & 1.1800 & 89.89 & 12.42 &  & 5 & 1.906 & 24.27 & 37.76 & 0.5049 & 1.0830 & 86.93 & 18.36 \\
 & 10 & 2.243 & 40.35 & 57.87 & 0.2683 & 0.5639 & 192.4 & 27.58 &  & 10 & 1.974 & 43.26 & 57.40 & 0.2715 & 0.4825 & 189.5 & 43.70 \\
 & 20 & 2.471 & 69.67 & 87.63 & 0.1428 & 0.2644 & 445.6 & 56.58 &  & 20 & 2.191 & 72.62 & 87.41 & 0.1434 & 0.2351 & 443.2 & 85.55 \\
\addlinespace[2pt]\multirow{4}{*}{EOS6} & 1 & 1.998 & 10.21 & 15.86 & 1.7130 & 2.9100 & 8.026 & -- & \multirow{4}{*}{RMF\_m55\_L50\_NY} & 1 & 2.143 & 13.95 & 17.81 & 1.4663 & 2.1518 & 0.4555 & $1.20\!\times\!10^{6}$ \\
 & 5 & 1.984 & 21.63 & 37.37 & 0.5127 & 1.2860 & 85.58 & 13.91 &  & 5 & 2.159 & 28.49 & 39.00 & 0.4818 & 0.9086 & 91.79 & 16.96 \\
 & 10 & 2.040 & 39.63 & 57.10 & 0.2737 & 0.5558 & 187.6 & 34.40 &  & 10 & 2.240 & 45.21 & 58.20 & 0.2661 & 0.4783 & 196.5 & 33.59 \\
 & 20 & 2.253 & 70.64 & 87.22 & 0.1438 & 0.2482 & 441.1 & 75.02 &  & 20 & 2.483 & 73.46 & 87.78 & 0.1445 & 0.2448 & 446.9 & 61.84 \\
\addlinespace[2pt]\multirow{4}{*}{EOS11} & 1 & 2.084 & 11.06 & 16.59 & 1.6060 & $2.4430^{\dagger}$ & 5.651 & -- & \multirow{4}{*}{RMF\_m55\_L60\_NY} & 1 & 2.140 & 13.94 & 17.83 & 1.4597 & 2.1528 & 0.4975 & $8.79\!\times\!10^{4}$ \\
 & 5 & 2.073 & 23.16 & 38.02 & 0.4995 & 1.1910 & 87.93 & 13.75 &  & 5 & 2.156 & 28.56 & 39.03 & 0.4814 & 0.9053 & 92.00 & 17.08 \\
 & 10 & 2.132 & 39.37 & 57.54 & 0.2706 & 0.5718 & 190.4 & 30.24 &  & 10 & 2.238 & 45.28 & 58.24 & 0.2658 & 0.4769 & 196.7 & 33.79 \\
 & 20 & 2.353 & 69.60 & 87.46 & 0.1433 & 0.2589 & 443.9 & 64.64 &  & 20 & 2.481 & 73.52 & 87.80 & 0.1445 & 0.2443 & 447.2 & 62.13 \\
\addlinespace[2pt]\multirow{4}{*}{EOS20} & 1 & 2.314 & 10.99 & 16.79 & 1.5790 & $2.5130^{\dagger}$ & 5.053 & -- & \multirow{4}{*}{RMF\_m60\_L50\_NY} & 1 & 2.007 & 13.84 & 17.51 & 1.4824 & 2.1792 & 1.984 & 699.4 \\
 & 5 & 2.282 & 20.21 & 38.14 & 0.4977 & 1.4710 & 88.68 & 8.630 &  & 5 & 2.030 & 29.47 & 38.80 & 0.4851 & 0.8459 & 90.96 & 21.09 \\
 & 10 & 2.318 & 35.32 & 57.62 & 0.2700 & 0.6915 & 191.0 & 20.27 &  & 10 & 2.113 & 46.80 & 58.05 & 0.2671 & 0.4433 & 195.5 & 41.76 \\
 & 20 & 2.531 & 65.08 & 87.48 & 0.1432 & 0.2958 & 444.0 & 46.59 &  & 20 & 2.349 & 74.91 & 87.72 & 0.1447 & 0.2317 & 446.2 & 74.95 \\
\addlinespace[2pt]\multirow{4}{*}{DD2Y} & 1 & 2.018 & 11.49 & 17.19 & 1.5238 & 2.6336 & 3.378 & 1143 & \multirow{4}{*}{RMF\_m60\_L60\_NY} & 1 & 2.002 & 13.84 & 17.57 & 1.4737 & 2.1786 & 1.775 & 791.7 \\
 & 5 & 2.021 & 25.19 & 38.52 & 0.4902 & 1.0515 & 90.04 & 16.71 &  & 5 & 2.026 & 29.53 & 38.83 & 0.4846 & 0.8430 & 91.24 & 21.27 \\
 & 10 & 2.092 & 43.41 & 57.90 & 0.2680 & 0.4924 & 192.6 & 37.63 &  & 10 & 2.109 & 46.90 & 58.09 & 0.2668 & 0.4415 & 195.6 & 42.13 \\
 & 20 & 2.318 & 72.12 & 87.63 & 0.1428 & 0.2438 & 445.6 & 72.17 &  & 20 & 2.345 & 75.00 & 87.75 & 0.1447 & 0.2311 & 446.3 & 75.51 \\
\addlinespace[2pt]\multirow{4}{*}{DDHdY4} & 1 & 2.033 & 11.40 & 16.92 & 1.5600 & 2.6490 & 4.381 & 1453 & \multirow{4}{*}{RMF\_m70\_L40\_N} & 1 & 2.314 & 10.38 & 16.76 & 1.5807 & 2.6503 & 4.808 & -- \\
 & 5 & 2.034 & 24.77 & 38.33 & 0.4939 & 1.0790 & 89.24 & 16.01 &  & 5 & 2.276 & 18.41 & 38.17 & 0.4968 & 1.6547 & 88.52 & 7.426 \\
 & 10 & 2.104 & 42.91 & 57.76 & 0.2690 & 0.5020 & 191.7 & 36.32 &  & 10 & 2.304 & 34.71 & 57.64 & 0.2699 & 0.7071 & 192.7 & 20.03 \\
 & 20 & 2.330 & 71.64 & 87.57 & 0.1429 & 0.2468 & 444.9 & 70.25 &  & 20 & 2.512 & 65.00 & 87.48 & 0.1453 & 0.2954 & 450.8 & 47.44 \\
\addlinespace[2pt]\multirow{4}{*}{GM1Y5} & 1 & 2.100 & 12.44 & 17.68 & 1.4550 & 2.4240 & 1.471 & 1185 & \multirow{4}{*}{RMF\_m70\_L50\_N} & 1 & 2.312 & 10.38 & 16.80 & 1.5757 & $2.5835^{\dagger}$ & 4.668 & -- \\
 & 5 & 2.107 & 26.50 & 38.93 & 0.4827 & 0.9954 & 91.67 & 16.20 &  & 5 & 2.273 & 18.44 & 38.23 & 0.4958 & 1.6511 & 88.71 & 7.471 \\
 & 10 & 2.181 & 43.23 & 58.23 & 0.2657 & 0.5044 & 194.6 & 33.41 &  & 10 & 2.302 & 34.80 & 57.67 & 0.2697 & 0.7042 & 192.9 & 20.16 \\
 & 20 & 2.416 & 72.36 & 87.87 & 0.1422 & 0.2472 & 447.9 & 64.79 &  & 20 & 2.510 & 65.12 & 87.51 & 0.1452 & 0.2945 & 443.9 & 47.70 \\
\addlinespace[2pt]\multirow{4}{*}{CMF-crust} & 1 & 2.054 & 12.01 & 17.54 & 1.4750 & 2.5180 & 1.750 & 1175 & \multirow{4}{*}{RMF\_m75\_L40\_N} & 1 & 2.107 & 10.34 & 16.40 & 1.6328 & $2.5608^{\dagger}$ & 6.324 & -- \\
 & 5 & 2.060 & 26.00 & 38.84 & 0.4843 & 1.0140 & 91.20 & 16.69 &  & 5 & 2.085 & 20.73 & 37.82 & 0.5037 & 1.3847 & 87.27 & 11.47 \\
 & 10 & 2.133 & 43.95 & 58.15 & 0.2662 & 0.4878 & 194.2 & 36.49 &  & 10 & 2.134 & 37.78 & 57.44 & 0.2712 & 0.6070 & 191.3 & 28.15 \\
 & 20 & 2.364 & 72.38 & 87.82 & 0.1423 & 0.2447 & 447.6 & 68.85 &  & 20 & 2.348 & 68.63 & 87.39 & 0.1456 & 0.2642 & 442.6 & 63.27 \\
\addlinespace[2pt]\multirow{4}{*}{SLy4} & 1 & 2.177 & 9.91 & 16.57 & 1.6082 & $2.7739^{\dagger}$ & 5.933 & -- & \multirow{4}{*}{RMF\_m75\_L50\_N} & 1 & 2.103 & 10.34 & 16.42 & 1.6290 & $2.7012^{\dagger}$ & 6.252 & -- \\
 & 5 & 2.140 & 18.61 & 37.97 & 0.5008 & 1.6077 & 87.90 & 9.046 &  & 5 & 2.081 & 20.80 & 37.85 & 0.5031 & 1.3779 & 87.50 & 11.57 \\
 & 10 & 2.173 & 37.04 & 57.55 & 0.2705 & 0.6291 & 190.4 & 25.96 &  & 10 & 2.131 & 37.89 & 57.47 & 0.2710 & 0.6039 & 191.5 & 28.40 \\
 & 20 & 2.382 & 68.34 & 87.47 & 0.1432 & 0.2676 & 443.9 & 60.36 &  & 20 & 2.345 & 68.77 & 87.42 & 0.1455 & 0.2632 & 442.8 & 63.74 \\
\addlinespace[2pt]\multirow{4}{*}{DD2\_ND} & 1 & 2.176 & 9.94 & 16.34 & 1.6410 & $2.7130^{\dagger}$ & 6.350 & -- &  &  &  &  &  &  &  &  &  \\
 & 5 & 2.141 & 18.82 & 37.77 & 0.5081 & 1.5850 & 87.10 & 9.197 &  &  &  &  &  &  &  &  &  \\
 & 10 & 2.178 & 37.25 & 57.40 & 0.2716 & 0.6248 & 189.6 & 26.04 &  &  &  &  &  &  &  &  &  \\
 & 20 & 2.389 & 68.30 & 87.41 & 0.1433 & 0.2682 & 443.1 & 59.81 &  &  &  &  &  &  &  &  &  \\
\bottomrule
\end{tabular}%
}
\end{table*}

%%%%%%%%%%%%%%%%%%%%%%%%%%%%%%%%%%%%%%%%%%%%%%%%%%%%%%%%%%%%%%%%%%%%%%%
\subsection{Accessibility of the tidal resonances}
\label{sec:res_access}

\begin{figure}[!htb]
  \centering
    \includegraphics[width=\linewidth]{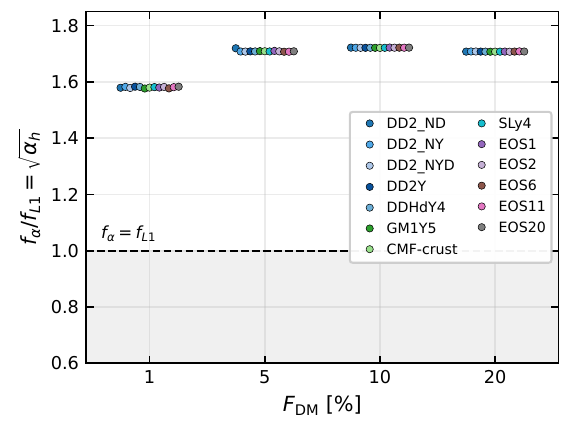}
\caption{Accessibility of the dark-matter-led halo resonance. For each of the
52 evaluated halo configurations at $1.4\,M_\odot$ (13 EoS $\times$
$F_{\rm DM}=1$--$20\%$, colors as in the legend), the equal-mass reference
ratio of the resonance frequency to the L1-overflow frequency,
$f_\alpha/f_{L1}=\sqrt{\alpha_h}$ with
$\alpha_h=(2\pi f_\alpha)^2R_{\rm DM}^3/GM$, is shown against the dark
fraction. Every configuration lies in the narrow band $1.58$--$1.72$ (mean
$1.68$ with population scatter $0.06$), above the accessibility boundary
(dashed line), so the spherical halo reaches L1 before the resonance under
this criterion.}
  \label{fig:res_access}
\end{figure}

\begin{table}[!htb]
  \caption{Resonance-accessibility criteria. $N_{\rm eval}$ is the number of
  cases tested and $N_{\rm pass}$ the number satisfying the indicated necessary
  condition. A zero therefore excludes a controlled resonance under the adopted
  criteria without requiring the remaining gates of
  \cref{eq:resonance_gates}. Equal-mass rows use
  $M=M'=1.4\,M_\odot$; core evaluations use a 12-km companion.}
  \label{tab:access_criteria}
  \footnotesize
  \begin{ruledtabular}
  \begin{tabular}{lcc}
  Criterion & $N_{\rm eval}$ & $N_{\rm pass}$ \\
  \colrule
  Halo, equal-mass L1 & 52 & 0 \\
  Halo, full-sequence Roche grid & $1{,}527\times1{,}301$ & 0 \\
  Core DM-led, equal-mass contact & 82 & 0 \\
  Core NM-led, equal-mass contact $+$ energy & 82 & 0 \\
  Core endpoints, all gates ($M'=2.3\,M_\odot$) & 84 & 0\footnote{Of
  the seven formally pre-contact NM-led endpoints, one lies marginally below
  the adopted energy threshold but fails the stationary-phase and point-mass
  conditions.} \\
  \end{tabular}
  \end{ruledtabular}
  \end{table}

\cref{fig:res_access} applies the accessibility conditions of
  \cref{eq:resonance_gates} to the halo sequences of
  Sec.~\ref{sec:halo_results}. Across the 52 evaluated configurations
  (13 EOS and four dark fractions), the resonance frequency exceeds the
  equal-mass L1 reference frequency by a mean factor $1.68$ with population
  scatter $0.06$. Thus, under this criterion, the halo reaches L1 before
  resonance. The weak variation of this ratio follows from the shared
  dynamical scale: both the DM-led frequency and the equal-mass L1 reference
  frequency scale approximately as $\sqrt{GM/R_{\rm DM}^{3}}$.
  Increasing $F_{\rm DM}$ lowers $f_\alpha$ (\cref{halo_freq}), but the
  accompanying halo expansion (\cref{mass_halo}) also lowers $f_{L1}$;
  none of the evaluated configurations thereby passes the pre-overflow
  condition.

  A broader evaluation uses all $1{,}527$ positive-frequency halo points
  from the full sampled sequences, including points outside the retained
  plotting sample. Testing each against $1{,}301$ companion masses from
  $1.0$ to $2.3\,M_\odot$ gives zero cases satisfying the adopted Roche-lobe
  criterion, with a maximum $R_L/R_{\rm DM}=0.906$
  (\cref{tab:access_criteria}). These tests use isolated spherical stellar
  radii and the corotating Roche geometry of \cref{eq:eggleton}.
  Nonsynchronous Roche models show that the critical geometry also depends
  on the rotation state~\cite{Sepinsky:2006if}. For an approximately
  irrotational inspiral, a quantitative overflow boundary therefore requires
  an appropriate binary treatment. The zero-pass result is consequently conditional on the adopted geometric
  prescriptions, rather than a dynamical prediction of tidal mass loss from
  the halo.

The compact core configurations fail for the complementary reason: their
  dark-matter-led frequencies are high (Sec.~\ref{sec:core_results}), so the
  resonance separation lies inside the summed material radii. For a
  $1.4{+}1.4\,M_\odot$ binary with a 12-km ordinary-neutron-star companion,
  the DM-led resonance separation lies $5.7$--$9.8$~km inside the contact
  separation in all 82 configurations, and no tested companion mass up to
  $2.3\,M_\odot$ changes this at the assumed companion radius. For a
  DM-admixed companion, a smaller outer radius would delay contact, while
  an extended halo would require its own overflow criterion.
  The NM-led branch admits a single formally pre-contact case, for which
  the leading linear formula would give
  $\Delta E_\alpha/|E_{\rm orb}|=1.4$, fourteen times the adopted $0.1$
  weak-response threshold and therefore a breakdown signal rather than a
  physical energy deposit. At $M'=2.3\,M_\odot$, 43 of the 82 NM-led
  resonances move formally outside contact, but their corresponding linear
  estimates remain $\Delta E_\alpha/|E_{\rm orb}|=0.54$--$0.99$.
  Among the maximum-mass endpoints, a single configuration falls marginally
  below the energy threshold while failing the stationary-phase and
  point-mass conditions
  ($\omega_\alpha t_{\rm res}\approx4.4$,
  $(R_{\rm NM}+R_{\rm comp})/D_{\rm res}=0.96$).
  Under the adopted binary prescriptions and within the nonrotating,
  circular-inspiral domain of \cref{eq:crown}, no configuration in the
  evaluated dataset satisfies all conditions for a controlled linear
  resonance. The halo and core DM-led modes fail the adopted Roche-overflow
  or contact criteria, whereas the remaining NM-led candidates fail the
  weak-response, stationary-phase, or point-mass conditions.
  This bounds the applicability of the resonant channel; it does not
  establish detectability or nondetectability through off-resonant tides,
  rotating or eccentric binaries, or post-contact dynamics.

%%%%%%%%%%%%%%%%%%%%%%%%%%%%%%%%%%%%%%%%%%%%%%%%%%%%%%%%%%%%%%%%%%%%%%%%%%%%%%%%%
\section{Discussion}
\label{sec:discussion}

\subsection{What the two branches measure}

  The two $f$-mode branches have complementary sensitivities. The NM-led
  frequency broadly tracks NM compactness and depends only weakly on
  $\lambda$ in our EOS1 core scan, whereas the DM-led frequency responds
  more strongly to the dark component's distribution and self-interaction
  through their effects on pressure support and the shared gravitational
  field. For core configurations at $1.4\,M_\odot$, the DM-led radiative
  damping time decreases by roughly three orders of magnitude between
  $F_{\rm DM}=1\%$ and $20\%$, approximately following the empirical trend
  $\tau_{\rm DM}^{1.4}\propto F_{\rm DM}^{-2}$.

  These complementary sensitivities matter because nuclear softening and
  a compact dark core can both reduce stellar radii and maximum masses;
  a frequency governed mainly by compactness need not distinguish their
  origins. The stronger fraction dependence of $\tau_{\rm DM}$ provides
  additional information, but not a unique measure of $F_{\rm DM}$.
  At $1.4\,M_\odot$, the damping time instead increases with fraction
  in halo configurations, and low-fraction halos overlap higher-fraction
  cores in damping time. Joint measurements of $f_{\rm NM}$,
  $f_{\rm DM}$, and $\tau_{\rm DM}$ could therefore help distinguish
  the dark distribution from nuclear-EOS effects. Whether these
  observables break parameter degeneracies requires a dedicated inference
  analysis, which is beyond the present survey.

%%%%%%%%%%%%%%%%%%%%%%%%%%%%%%%%%%%%%%%%%%%%%%%%%%%%%%%%%%%%%%%%%%%%%%%%%%%%%%%%%%%%%%%
\subsection{Detectability}

  Assessing whether the predicted modes are observable with the
  Einstein Telescope or Cosmic Explorer requires an excitation model
  and a detector-response analysis~\cite{Punturo_2010,
  LIGOScientific:2016wof,Kalogera:2021bya}. Koehn et al.\ have studied
  Einstein Telescope constraints on fermionic DM through its effect
  on equilibrium tidal deformability~\cite{Koehn:2024gal}.
  Our tidal analysis instead tests the accessibility of mode resonances
  and does not provide a detector forecast. The core-induced increase
  in the NM-led frequency could bias a single-fluid interpretation of
  the nuclear EOS if the dark component were neglected, but its
  measurability cannot be established from the frequency shift alone.

  For the DM-led core branch, increasing the dark fraction shortens the
  radiative lifetime but also raises the frequency; neither change alone
  establishes improved detectability. For a narrowband ringdown observed
  over its full decay, the matched-filter signal-to-noise ratio scales as
  $\mathrm{SNR}\propto E_{\rm GW}^{1/2}/[Df\sqrt{S_n(f)}]$, up to angular
  factors, where $E_{\rm GW}$ is the radiated energy, $D$ the source
  distance, and $S_n$ the one-sided detector noise spectral
  density~\cite{Sutton:2013ooa}. A longer lifetime reduces the instantaneous
  radiative power at fixed mode energy but extends the signal; its effect
  on detectability depends on the observing duration and competing
  dissipation. The tabulated frequencies and damping times therefore do
  not, by themselves, determine the relative detection reach of the
  branches. An excitation model and detector-level analysis are required.

  For resonant inspiral signals, Yu and Weinberg found hyperonic
  $g$-mode phase shifts of order $10^{-3}\,\mathrm{rad}$ difficult to
  detect; their favorable stacking prospects concern the larger
  muonic-mode signals~\cite{Yu:2017cxe}. A study using full-GR
  frozen-composition $g_1$ modes reported
  $|\Delta\Phi_{g_1}|\leq1.410\times10^{-3}\,\mathrm{rad}$,
  about 21 times below its adopted $0.03$-rad favorable-event ET
  benchmark~\cite{Thakur:2026cac}. This benchmark is not a universal
  detection threshold. A detector-level analysis of our DM-led modes
  with ET or CE lies outside the present survey
  \cite{ET:2025xjr,Evans:2021gyd}.

  The large-fraction halo modes reach the hundred-hertz range, but this
  does not make their isolated resonances accessible. Halo expansion
  lowers both the mode frequency and the Roche-overflow frequency through
  their shared dynamical scale. In the evaluated halo sample, the adopted
  Roche criterion is reached before resonance
  (\cref{sec:res_access}), so \cref{eq:crown} provides no controlled
  phase estimate for these cases. This restricts the clean-resonance
  channel in the adopted nonrotating, circular-binary model, not all
  possible GW signatures. Off-resonant tides, rotation, eccentricity,
  tidal mass loss from the halo, and post-contact excitation remain
  outside the present calculation. Numerical-relativity simulations
  have explored mergers of neutron stars admixed with fermionic
  DM~\cite{Giangrandi:2025rko}. Simulations of rapidly and retrogradely
  spinning neutron-star binaries have also resolved $f$-mode tidal
  resonance~\cite{Kuan:2024jnw}, while a dedicated Bayesian study has
  assessed tidal-resonance detectability with the Einstein
  Telescope~\cite{Gittins:2026ntx}. Extending the present bosonic
  two-fluid analysis to rotating binaries and merger dynamics remains
  for future work.

%%%%%%%%%%%%%%%%%%%%%%%%%%%%%%%%%%%%%%%%%%%%%%%%%%%%%%%%%%%%%%%%%%%%%%%%%%%%%%%%
\subsection{Relation to previous work}

  The core--halo structural trends are qualitatively consistent with
  earlier studies of repulsively self-interacting bosonic DM
  \cite{Karkevandi:2021ygv,Karkevandi:2024vov}. Component-associated mode
  families and low-frequency halo oscillations were already identified
  in two-fluid Cowling calculations
  \cite{Sotani:2025hzb,Sotani:2025lzy}. Kumar et al.\ developed the
  full-GR polar framework and computed real-frequency spectra for
  mirror-DM configurations, without extracting radiative damping
  times~\cite{Kumar:2026tgi}. Zheng et al.\ obtained complex frequencies
  and damping times across core and halo configurations for
  self-interacting fermionic DM with one microscopic nuclear
  EOS~\cite{Zheng:2026fwf}. The present study complements these
  developments with a full-GR complex-mode survey of bosonic DM across
  21 nuclear EOS, examining the dependence on nuclear composition,
  dark fraction, and core--halo geometry.

  Our strongest dark-core endpoint upturns are qualitatively consistent
  with the high-mass suppression of outgoing-wave coupling identified
  by Zheng et al., rather than their low-mass matter-quadrupole
  cancellation~\cite{Zheng:2026fwf}. In our pronounced upturns, $R_Q$
  increases toward the endpoint, while the leading-quadrupole diagnostic
  underpredicts the strongest damping-time growth. The ensemble
  establishes how the occurrence and strength of this weakly radiating
  behavior vary with the nuclear EOS and dark fraction. We additionally
  test whether the computed modes can undergo a controlled linear
  tidal resonance before contact or Roche overflow, finding no
  configuration that satisfies all required conditions under the
  adopted binary prescriptions (\cref{sec:res_access}).

%%%%%%%%%%%%%%%%%%%%%%%%%%%%%%%%%%%%%%%%%%%%%%%%%%%%%%%%%%%%%%%%%%
 \subsection{Limitations}

  We consider nonrotating, unmagnetized stars composed of cold barotropic
  NM and a repulsively self-interacting bosonic condensate; the dark
  sector is fixed to a bosonic condensate throughout, and we do not
  attempt the fermionic-vs-bosonic comparison carried out
  elsewhere~\cite{Arvikar:2025dwl}. The two components interact only
  through gravity; direct NM--DM interactions, finite-temperature
  effects, crustal elasticity, and NM superfluidity are omitted.
  Their effects on the mode spectrum are therefore not quantified here.
  The main survey fixes $\lambda=\pi$, while the self-coupling scan is
  restricted to EOS1 core configurations; the corresponding dependence
  of halo modes remains unexplored.

  The calculated damping times describe GW emission within the
  ideal-fluid QNM model. Shear and bulk viscosity, together with possible
  dark-sector dissipation, can modify the physical decay time.
  Establishing the dominant damping mechanism requires evaluating these
  processes using the relevant microphysics and each mode's fluid
  motion. The present calculation therefore does not establish a
  dark-fraction threshold above which the radiative damping time equals
  the total physical lifetime.

  The dark fraction is prescribed rather than derived from a formation
  or capture history. Percent-level admixtures require a scenario beyond
  standard ambient capture in the solar neighbourhood
  \cite{Bramante:2023djs}. Finally, although the isolated stellar modes
  are calculated in full GR, the tidal-dephasing estimate uses
  leading-order source and orbital matching,
  $\kappa_{\rm tide}=\kappa_{\rm orb}=1$, subject to the applicability
  conditions of \cref{eq:resonance_gates} and the accessibility tests
  of \cref{sec:res_access}. The Roche criterion is a quasi-equilibrium
  prescription, not a dynamical calculation of halo mass transfer, and
  the phase estimate is not a complete binary waveform.

%%%%%%%%%%%%%%%%%%%%%%%%%%%%%%%%%%%%%%%%%%%%%%%%%%%%%%%%%%%%%%%%%%%%%%%%%%%%%%%%%%%%
\section{Conclusion}
  \label{conc}

  We have computed quadrupolar $f$-mode frequencies and radiative damping
  times of bosonic DM-admixed neutron stars in full GR, using 21 nuclear
  EOS spanning nucleonic, hyperonic, $\Delta$-admixed, and combined
  hyperon--$\Delta$ compositions. The survey covers
  $F_{\rm DM}=1\%,5\%,10\%$, and $20\%$ in both core and halo geometries.
  The spectrum contains two principal branches: the NM-led branch
  connects to the ordinary single-fluid $f$ mode, while the additional
  DM-led branch reflects the independent motion of the dark component.

  For the adopted core configurations, increasing the dark fraction
  compresses the NM component, generally raises the mode frequencies,
  and reduces the maximum mass. The most pronounced spectral change is
  in the DM-led radiative lifetime. At $1.4\,M_\odot$, it decreases from
  $395$--$1317\,\mathrm{s}$ at $F_{\rm DM}=1\%$ to
  $0.7$--$1.6\,\mathrm{s}$ at $20\%$, approximately following an empirical
  $F_{\rm DM}^{-2}$ dependence. NM-led damping varies much less strongly.
  Selected high-mass core sequences also develop pronounced damping-time
  upturns, showing that radiative efficiency cannot be inferred from
  frequency alone.

  Extended halos produce the opposite DM-led trend: increasing the dark
  fraction lowers the frequency and lengthens the radiative lifetime,
  while separating the NM surface from the outer dark radius. The
  canonical $20\%$ halo models reach frequencies near
  $0.14\,\mathrm{kHz}$ and damping times of several hundred seconds.
  At large dark fractions, the halo maximum masses increase, contrasting
  with the core sequences.

  The EOS1 core self-coupling scan shows greater sensitivity of the
  DM-led branch than of the NM-led branch, and the representative
  octupolar calculation retains the two-branch structure. These results
  motivate joint frequency and damping-time diagnostics. Both branches,
  however, depend on the coupled fluid and spacetime structure;
  demonstrating independent constraints on the nuclear EOS and dark
  sector requires a dedicated inference analysis.

  For the evaluated nonrotating circular binaries, no configuration
  satisfies all conditions for a controlled linear tidal resonance under
  the adopted applicability criteria. Halo and core DM-led resonances
  are pre-empted by Roche overflow or contact, while the remaining
  NM-led candidates fail the energy, timescale, or point-mass conditions.
  This bounds the tested resonant-inspiral channel, not general
  detectability. Excitation and detector modelling, together with
  rotating and finite-temperature extensions, are needed to assess
  other observational channels. Tables I–II supply the isolated-mode inputs for that modelling, spanning roughly $0.14$–$4.9\,\mathrm{kHz}$ across the evaluated configurations.

\begin{acknowledgments}
We are grateful to J\"{u}rgen Schaffner-Bielich for helpful discussions, valuable suggestions, and comments on the manuscript.
P.~T. and Y.~L. are supported by a National Research Foundation of Korea (NRF) grant funded by the Korean government (MSIT) (No.~RS-2024-00457037).
P.~T. was also supported in part by the Yonsei University Research Fund (Yonsei University Frontier Fellowship for Postdoctoral Researchers) in 2025.
Y.~L. acknowledges support from the Global–Learning \& Academic Research Institution for Master’s and PhD Students, and Postdocs (LAMP) Program through a National Research Foundation of Korea (NRF) grant funded by the Ministry of Education (No.~RS-2024-00442483).
I. A. R. gratefully acknowledges support from the Deutsche Forschungsgemeinschaft (DFG, German Research Foundation) – Project Number 579861443 and also in part by the Alexander von Humboldt Foundation through a Humboldt Research Fellowship.

\end{acknowledgments}

\appendix
\section{Resonant tidal dephasing: derivation}
\label{app:tidal}

\subsection{Relativistic tidal-resonance reduction}

  Here $\omega$ denotes physical angular frequency, obtained by
  multiplying the solver's inverse-length frequency by $c$.
  With the adopted $e^{i\omega t}$ convention, a damped QNM has
  $\omega_{\rm QNM}=\omega_\alpha+i/\tau_\alpha$, where
  $\omega_\alpha=2\pi f_\alpha$ and $\tau_\alpha$ is the radiative
  amplitude-damping time.

  For the ideal-fluid star, flux conservation and time reversal imply
  a conjugate pole--zero pair in the exterior reflection coefficient
  $\mathcal R=A_{\rm out}/A_{\rm in}$~\cite{Miao:2025utd}.
  Near an isolated pole,
  \begin{equation}
  \begin{aligned}
  \mathcal R(\omega)
   &=\mathcal R_{0,\alpha}(\omega)
   \frac{\omega-\omega_{\rm QNM}^{*}}
        {\omega-\omega_{\rm QNM}},\\
  \mathcal B_\alpha
   &=|\mathcal R_{0,\alpha}(\omega_{\rm QNM})|.
  \end{aligned}
  \end{equation}
  The regular factor $\mathcal R_{0,\alpha}$ has unit modulus on the
  real-frequency axis. Taking the residue therefore gives
  $|\operatorname*{Res}\mathcal R|=2\mathcal B_\alpha/\tau_\alpha$,
  as in \cref{eq:gr_reciprocity}.

  For a quadrupolar tide with driving frequency $\sigma=2\Omega$,
  stationary phase supplies a squared-amplitude factor
  $2\pi/\dot{\sigma}_{\rm res}$. Using the single-harmonic Zerilli
  luminosity
  $\mathcal F_{22}=(3c^3/8\pi G)|\dot Z_{22}|^2$
  and integrating the freely decaying response gives
  \begin{equation}
  \Delta E_{22}
   =\frac{3c^3}{2G}
   \frac{\mathcal B_\alpha^2\omega_\alpha^2|A_0|^2}
        {\dot{\sigma}_{\rm res}\tau_\alpha},
  \end{equation}
  where $A_0$ is the incident time-domain amplitude in this
  normalization. Matching the leading quadrupolar companion field
  to the same Zerilli basis gives
  \begin{equation}
  |A_0|^2
   =\kappa_{\rm tide}\frac{15\pi}{8}
   \frac{G^2M'^2c^2}{D_{\rm res}^6\omega_\alpha^6}.
  \end{equation}
  Here $\kappa_{\rm tide}$ accounts for corrections to the leading
  source matching; its adopted value is unity. It must be defined
  consistently with the asymptotic conventions used for $\mathcal R$.

  Including the equal contribution from the conjugate $m=-2$
  harmonic and using $\dot{\sigma}_{\rm res}=2\dot{\Omega}_{\rm res}$
  gives the total deposited energy in one star,
  \begin{equation}
  \Delta E_\alpha
   =\mathcal B_\alpha^2\kappa_{\rm tide}
   \frac{45\pi}{16}
   \frac{Gc^5M'^2}
   {D_{\rm res}^6\dot{\Omega}_{\rm res}\omega_\alpha^4\tau_\alpha}.
  \label{eq:deltaE_res}
  \end{equation}

  For a leading-order circular point-mass inspiral~\cite{Peters:1964zz},
  \begin{equation}
  \begin{aligned}
  \dot{\Omega}_{\rm res}
   &=\frac{96}{5}\frac{G^{5/3}}{c^5}
     \mu M_t^{2/3}\Omega_{\rm res}^{11/3},\\
  \mathcal F_{\rm GW}
   &=\frac{32}{5}\frac{G^{7/3}}{c^5}
     \mu^2M_t^{4/3}\Omega_{\rm res}^{10/3},
  \end{aligned}
  \end{equation}
  where $\mu=MM'/M_t$. Combining these expressions with
  $D_{\rm res}^3=GM_t/\Omega_{\rm res}^2$,
  $\omega_\alpha=2\Omega_{\rm res}$, and
  \begin{equation}
  \delta\Phi_\alpha
   =-\frac{2\Omega_{\rm res}\Delta E_\alpha}{\mathcal F_{\rm GW}},
  \qquad
  \Delta\Phi_\alpha=|\delta\Phi_\alpha|,
  \end{equation}
  recovers \cref{eq:crown}, including the coefficient
  $375/(131072\pi^5)$. At this orbital order
  $\kappa_{\rm orb}=1$. The stellar response is fully relativistic,
  while the binary source and orbital evolution enter at the stated
  leading order; the residue alone does not specify an arbitrarily
  normalized external forcing.

  As an independent leading-quadrupole check, the same complex-harmonic
  convention gives $W_{22}^2=3\pi/10$ and
  \begin{equation}
  \Delta E_\alpha^{\rm quad}
   =\frac{\pi}{2\dot{\Omega}_{\rm res}}
   \left(\frac{GM'W_{22}}{D_{\rm res}^3}\right)^2
   \frac{|Q_\alpha|^2}{J_\alpha}.
  \end{equation}
  Eliminating $|Q_\alpha|^2/J_\alpha$ with \cref{eq:duality}
  recovers the same phase coefficient with
  $\mathcal C_\alpha=1$ and $\tau_\alpha^{\rm quad}$ replacing
  $\tau_\alpha$. This is a diagnostic cross-check, not an
  identification of the fluid-only inertia with an exact GR QNM norm.
  The reduction applies only under the conditions of
  \cref{eq:resonance_gates}.

%%%%%%%%%%%%%%%%%%%%%%%%%%%%%%%%%%%%%
\section{Supplementary results and numerical diagnostics}
\label{app:result}

\subsection{Tidal-resonance diagnostics and validity conditions}

  Full-GR residue calculations give $\mathcal B_\alpha\simeq1.00008$
  for the reference core NM-led mode and $0.9999991$ for the halo
  DM-led mode. Residue and tidal-source matching must share the same
  tortoise-coordinate origin and asymptotic basis; their combined
  factor $\mathcal C_\alpha$ enters the phase estimate. Using the real parts of the corresponding QNM frequencies, the
  leading-quadrupole diagnostic gives
  $\tau_\alpha^{\rm quad}/\tau_\alpha=1.0439$ and $0.9095$:
  differences of $4.4\%$ and $9.1\%$, respectively.
  Surface terms are evaluated separately; their combined magnitude
  is below $3\times10^{-5}$ of the halo bulk overlap, negligible
  relative to this percent-level comparison.

  Beyond \cref{eq:resonance_gates}, the leading tidal and orbital
  reduction requires
  \begin{equation}
  \frac{R_{{\rm out},i}}{D_{\rm res}}\ll1,
  \qquad
  \frac{GM_t}{c^2D_{\rm res}}\ll1,
  \end{equation}
  and a slowly varying forcing amplitude across $t_{\rm res}$.
  Linear response and negligible feedback on the orbital frequency
  sweep require, respectively,
  \begin{equation}
  \begin{aligned}
  \Delta E_\alpha &\ll E_{{\rm nonlinear},\alpha},\\
  \Delta E_\alpha &\ll \mathcal F_{\rm GW}t_{\rm res}.
  \end{aligned}
  \end{equation}
  The mode- and configuration-dependent nonlinear threshold is not
  computed here. The proxy $GM_{\rm D}^2/R_{\rm D}$ includes only
  DM self-gravity, omitting NM binding, dark self-interaction, and
  nonlinear fluid dynamics. Tidal mass loss requires a separate
  dynamical treatment.

\subsection{Representative core-mode eigenfunctions}
  \label{app:core_eigenfunctions}

\begin{figure*}[p]
    \centering
    \includegraphics[width=0.85\linewidth]
      {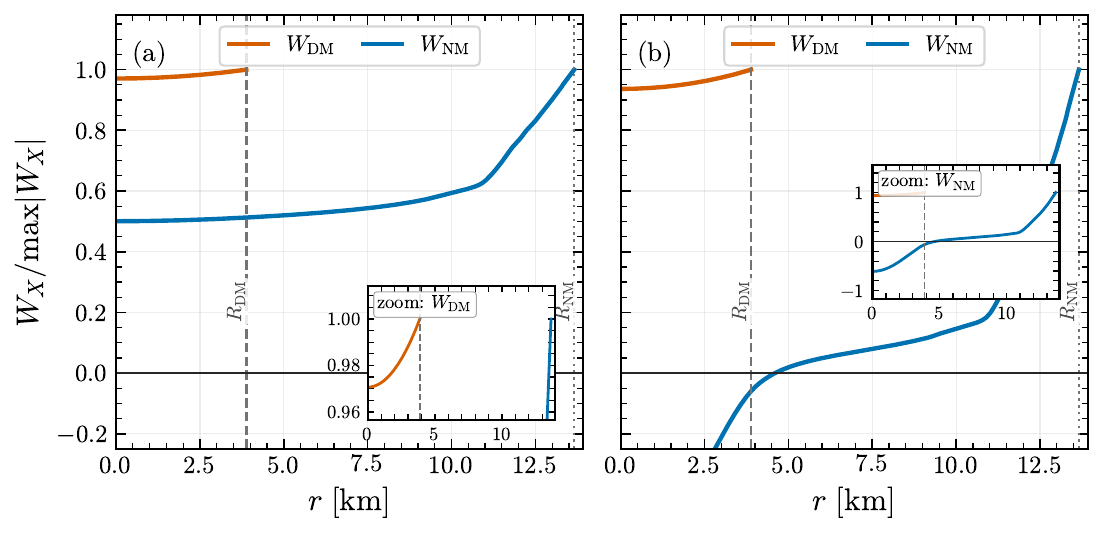}
    \caption{Normalized radial-displacement eigenfunctions of the two
    fundamental quadrupolar ($\ell=2$) $f$-modes of a
    $0.501\,M_\odot$ dark-matter-core configuration: EOS1 normal matter
    admixed with bosonic dark matter ($m_\chi=500\,\mathrm{MeV}$,
    repulsive self-interaction $\lambda=\pi$) at dark-matter fraction
    $F_{\rm DM}=1\%$. Panel~(a): the normal-matter-led mode,
    $f^{\rm O}=1.245\,\mathrm{kHz}$, with node counts
    $(n_{\rm NM},n_{\rm DM})=(0,0)$. Panel~(b): the dark-matter-led
    mode, $f^{\rm I}=2.150\,\mathrm{kHz}$, with
    $(n_{\rm NM},n_{\rm DM})=(1,0)$. Orange and blue curves show
    the dark- and normal-fluid displacements. Each curve is divided
    by its value where $|W_X|$ is largest. This independent
    normalization displays shapes and nodes, but not the relative
    amplitudes or phase relation between the two fluids.
    The dark matter is the inner fluid: the dashed vertical line
    marks $R_{\rm DM}=3.88\,\mathrm{km}$, and the dotted vertical
    line marks the outer normal-matter surface at
    $R_{\rm NM}=13.66\,\mathrm{km}$. Insets enlarge the DM profile
    in panel~(a) and the NM profile in panel~(b).}
    \label{fig:core_eigenfunctions}
  \end{figure*}

    \cref{fig:core_eigenfunctions}(a) shows the NM-led mode
  $f^{\rm O}$, with nodeless motion in both components.
  The NM displacement extends through the normal-matter envelope,
  while the dark displacement terminates at the inner surface
  $R_{\rm DM}$. The dark core nevertheless participates dynamically:
  it is not simply a fixed contribution to the background gravitational
  field. This branch connects continuously to the ordinary NM
  $f$-mode as the dark fraction vanishes, while the finite dark
  component introduces an accompanying fluid response through the
  common metric perturbations.

  Panel~(b) shows the distinct spatial structure of the DM-led mode
  $f^{\rm I}$. The dark displacement remains nodeless, but the NM
  displacement changes sign just outside $R_{\rm DM}$, as highlighted
  by the inset. This node separates NM regions with opposite radial
  oscillation phases; it does not contradict the fundamental character
  of the branch, whose dominant dark-fluid displacement has no node.
  The NM response continues beyond the dark surface, illustrating
  that the dark-core oscillation is coupled to the surrounding
  normal matter rather than confined to an isolated bosonic object.
  Its higher frequency is consistent with the shorter dynamical
  timescale of the compact inner region, where dark-fluid pressure
  and the gravitational field generated by both components provide
  the restoring dynamics. The contrast between panels~(a) and~(b)
  therefore concerns the organization of motion throughout the
  two-fluid star, not merely a shift in the frequency of the same
  NM oscillation.

\begin{figure*}[p]
    \centering
    \includegraphics[width=\linewidth]
      {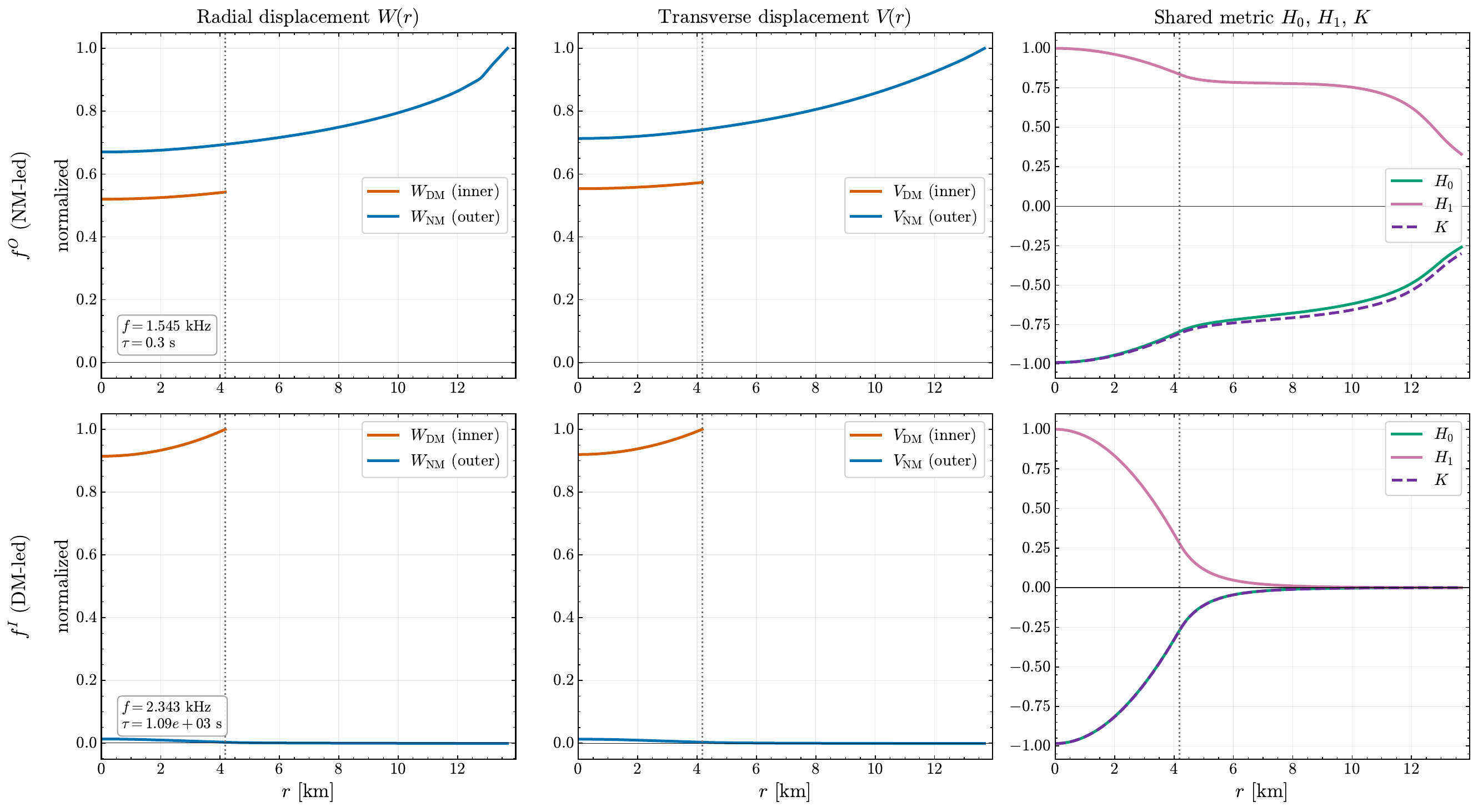}
    \caption{Fluid-displacement and metric profiles of the two
    quadrupolar ($\ell=2$) modes of a representative
    $M\simeq1.40\,M_\odot$ core configuration: EOS1 NM and bosonic
    DM with $m_\chi=500\,\mathrm{MeV}$, $\lambda=\pi$, and
    $F_{\rm DM}=1\%$. The top row shows the NM-led mode
    ($f^{\rm O}\simeq1.545\,\mathrm{kHz}$,
    $\tau\simeq0.30\,\mathrm{s}$); the bottom row shows the
    DM-led mode ($f^{\rm I}\simeq2.343\,\mathrm{kHz}$,
    $\tau\simeq1.1\times10^3\,\mathrm{s}$).
    Profiles are evaluated at the displayed real-frequency
    estimates. Left, middle, and right columns show $W$, $V$,
    and the metric amplitudes $(H_0,H_1,K)$, respectively.
    Orange and blue curves denote the dark and normal fluids.
    All curves within a panel share one signed normalization
    factor; different panels are normalized independently.
    The dotted vertical line marks the dark-fluid surface.}
    \label{fig:core_edge}
  \end{figure*}

\cref{fig:core_edge} extends the comparison to a
  $1.4\,M_\odot$ core configuration and shows the fluid motion
  together with its spacetime response. In the top row, the NM
  component dominates both the radial and transverse displacements,
  while the dark core also participates within its own surface.
  The top-right metric profiles remain appreciable throughout the
  NM envelope and at its outer boundary, where the interior
  perturbation is matched to the outgoing Zerilli solution.
  The NM-led mode consequently retains a radiative damping time
  near $0.30\,\mathrm{s}$, characteristic of the ordinary
  neutron-star $f$-mode.

  The bottom row shows a different distribution of the perturbation.
  The radial and transverse motion is dominated by the compact
  dark component, with a much smaller NM response. The metric
  perturbations are largest in the inner region and decrease
  strongly across the NM envelope, becoming small at the outer
  surface. Thus, a mode can involve substantial internal fluid
  motion while coupling only weakly to outgoing gravitational
  waves. The radiative calculation quantifies this contrast:
  the DM-led lifetime is approximately
  $1.1\times10^3\,\mathrm{s}$, more than three orders of magnitude
  longer than that of the NM-led mode. For this DM-led solution,
  the opposing NM and DM contributions partially cancel in the
  leading matter-quadrupole diagnostic. The full lifetime,
  however, follows from the complete fluid--metric eigenmode and
  its exterior-wave matching. The shared metric therefore both
  couples the fluid motions and connects them to the radiation
  channel; these dynamical spacetime perturbations are omitted
  in the Cowling approximation.

\subsection{Dark-matter self-interaction}

  \begin{figure}[!htb]
    \centering
    \includegraphics[width=\linewidth]{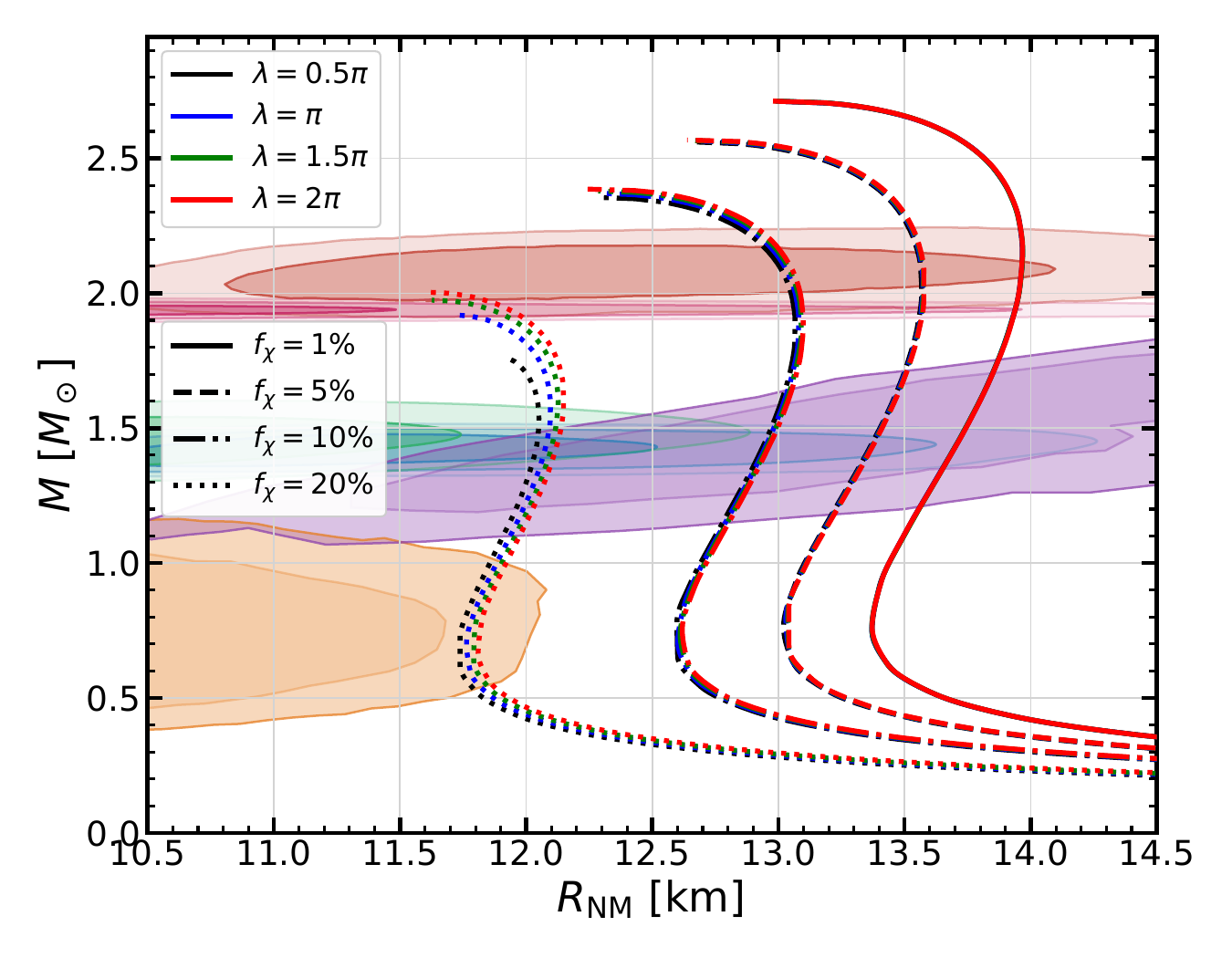}
    \caption{Total gravitational mass versus outer normal-matter
    radius $R_{\rm NM}$ for EOS1 with bosonic dark cores
    ($m_\chi=500\,\mathrm{MeV}$). Colors denote the self-coupling
    $\lambda=0.5\pi,\,\pi,\,1.5\pi$, and $2\pi$.
    Line styles distinguish the dark-matter fraction
    $F_{\rm DM}$ (denoted $f_\chi$ in the legend):
    $1\%$ (solid), $5\%$ (dashed), $10\%$ (dash-dotted), and
    $20\%$ (dotted). The shaded observational regions are
    the same as in \cref{mass_radius}.}
    \label{mr_diffself}
  \end{figure}

  \cref{mr_diffself} shows that the sensitivity of the maximum
  mass to the dark self-coupling grows with the dark fraction.
  The four coupling choices give nearly coincident curves at
  $F_{\rm DM}=1\%$, whereas at $20\%$, increasing $\lambda$
  from $0.5\pi$ to $2\pi$ raises $M_{\max}$ from approximately
  $1.76$ to $2.00\,M_\odot$, recovering about $0.25\,M_\odot$.
  The canonical radius is much less sensitive: at
  $1.4\,M_\odot$, the same coupling variation increases
  $R_{\rm NM}$ by only $0.08\,\mathrm{km}$ even at $20\%$.
  Increasing the dark fraction nevertheless reduces both
  $R_{1.4}$ and $M_{\max}$ throughout this scan.

  This contrast follows from the repulsive bosonic EOS:
  increasing $\lambda$ raises the pressure at fixed dark
  energy density (\cref{eq:bosonic_dm_eos}). At fixed total
  mass and dark fraction, the stronger repulsion supports
  a more extended dark core, redistributing its gravitational
  influence on the normal fluid. This support acts within
  the dark component; the normal matter responds through
  gravity, not direct pressure coupling. Consequently,
  stronger self-interaction partially offsets the reduction
  of the maximum mass caused by the dark core, with the
  largest effect occurring when the dark component carries
  a substantial fraction of the mass. This maximum-mass
  trend agrees with the bosonic survey of
  Karkevandi \textit{et al.}~\cite{Karkevandi:2021ygv}.

\begin{figure*}[!tb]
    \centering
    \includegraphics[width=0.85\linewidth]{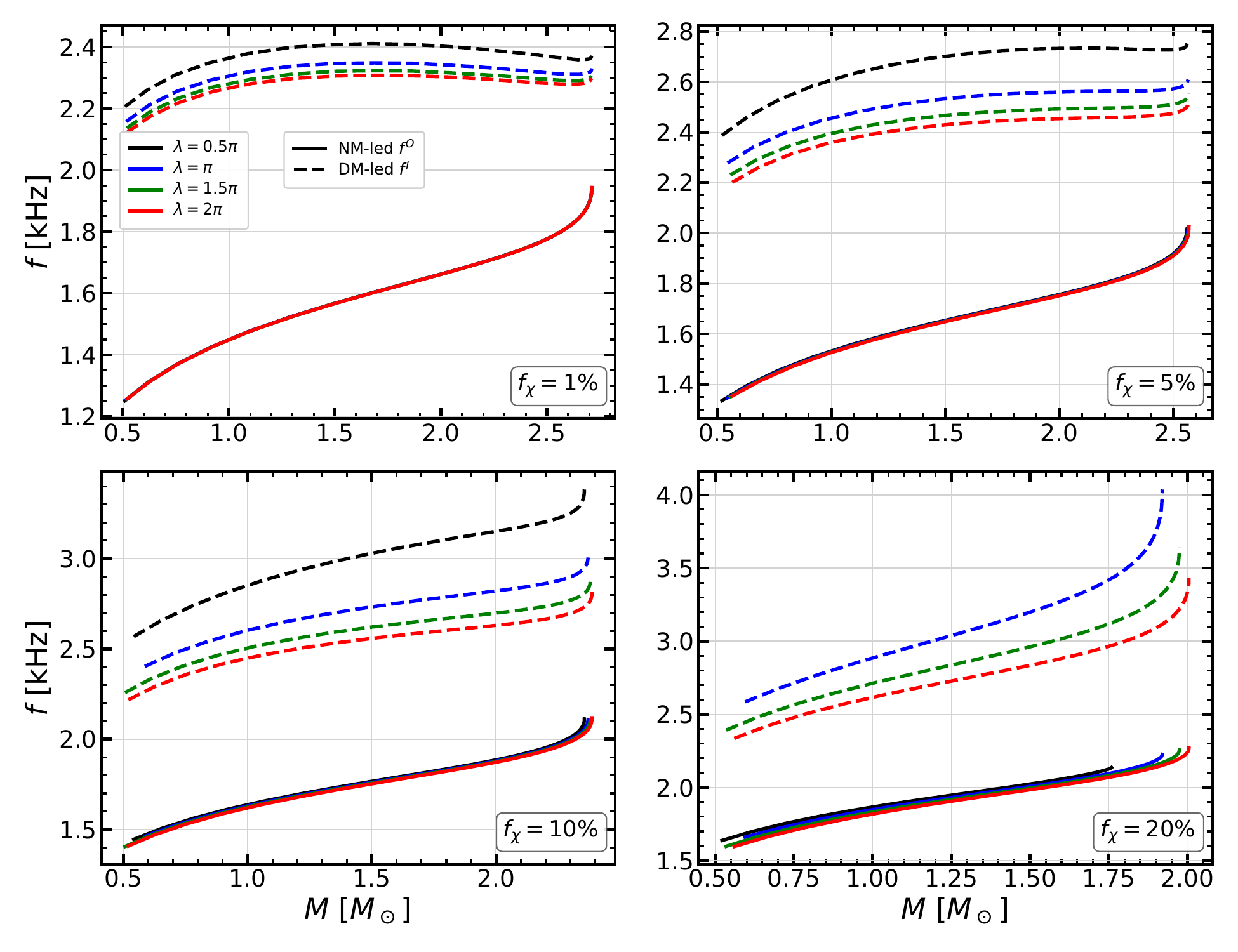}
    \caption{Full-GR quadrupolar $f$-mode frequencies versus total
    gravitational mass for EOS1 with bosonic dark cores
    ($m_\chi=500\,\mathrm{MeV}$). Colors denote
    $\lambda=0.5\pi,\,\pi,\,1.5\pi$, and $2\pi$.
    The panels correspond, in reading order, to
    $F_{\rm DM}=1\%,\,5\%,\,10\%$, and $20\%$
    (denoted $f_\chi$ in the panels). Solid and dashed curves
    represent the NM-led branch $f^{\rm O}$ and DM-led branch
    $f^{\rm I}$, respectively. No resolved DM-led sequence is
    available for $\lambda=0.5\pi$ at $F_{\rm DM}=20\%$
    in the plotted dataset.}
    \label{dm_self_fmode}
  \end{figure*}

  \cref{dm_self_fmode} shows that the DM-led frequency is
  considerably more sensitive to the dark self-coupling than
  the NM-led frequency. At fixed mass and dark fraction,
  increasing $\lambda$ lowers the resolved DM-led frequencies.
  For example, at $1.4\,M_\odot$ and $F_{\rm DM}=10\%$,
  increasing $\lambda$ from $0.5\pi$ to $2\pi$ reduces
  $f^{\rm I}$ from $3.00$ to $2.54\,\mathrm{kHz}$, while
  $f^{\rm O}$ changes by only $0.014\,\mathrm{kHz}$.
  At this canonical mass, the NM-led spread remains below
  $0.04\,\mathrm{kHz}$ across all four dark fractions.

  Although stronger repulsion raises the dark pressure at
  fixed density, it also expands the dark core and changes
  its equilibrium profile (\cref{mr_diffself}); the DM-led
  frequency decreases along these readjusted configurations.
  Its restoring dynamics involve the dark-fluid pressure
  and the gravitational field generated by both components,
  rather than the dark mean density alone. The smaller
  NM-led shift is consistent with the weak response of
  $R_{\rm NM}$ to $\lambda$ at fixed mass. Thus, this scan
  identifies the DM-led branch as the more sensitive probe
  of the dark self-interaction, without implying that the
  NM-led mode is independent of dark-matter admixture.

\begin{figure*}[!tb]
    \centering
    \includegraphics[width=0.85\linewidth]{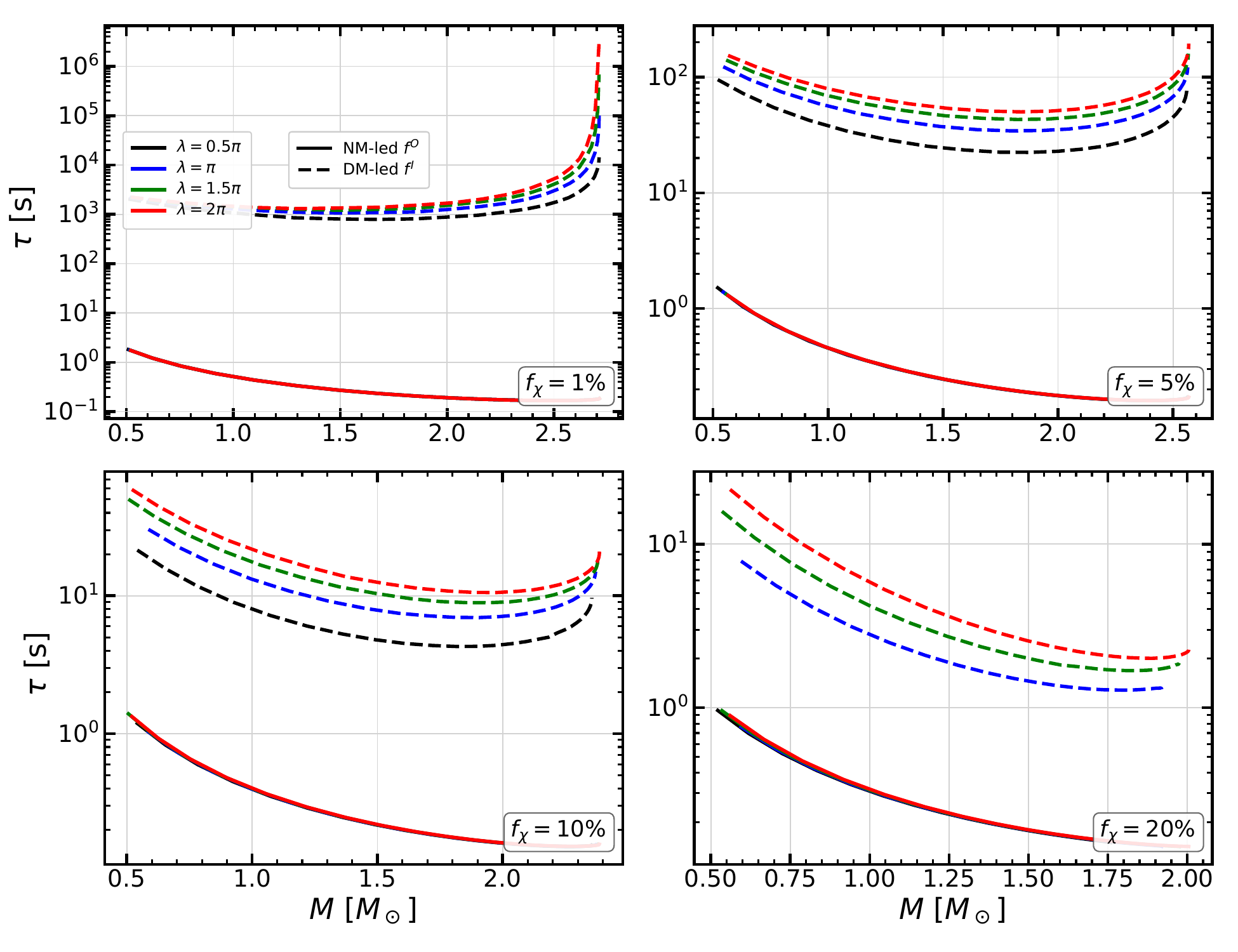}
    \caption{Full-GR radiative damping times of the quadrupolar
    $f$-modes versus total gravitational mass for EOS1 with
    bosonic dark cores ($m_\chi=500\,\mathrm{MeV}$).
    Colors denote $\lambda=0.5\pi,\,\pi,\,1.5\pi$, and $2\pi$.
    The panels correspond, in reading order, to
    $F_{\rm DM}=1\%,\,5\%,\,10\%$, and $20\%$
    (denoted $f_\chi$ in the panels). Solid and dashed curves
    represent the NM-led and DM-led branches, respectively;
    all damping-time axes are logarithmic. No resolved DM-led
    sequence is available for $\lambda=0.5\pi$ at
    $F_{\rm DM}=20\%$ in the plotted dataset.}
    \label{dm_self_tau}
  \end{figure*}

  \cref{dm_self_tau} shows that increasing $\lambda$ generally
  lengthens the DM-led damping time, reversing the frequency
  ordering in \cref{dm_self_fmode}. At $1.4\,M_\odot$ and
  $F_{\rm DM}=10\%$, increasing $\lambda$ from $0.5\pi$ to
  $2\pi$ raises $\tau_{\rm DM}$ by a factor of approximately
  $2.6$. The canonical NM-led damping time is much less
  sensitive, varying by less than $1.2\%$ across the couplings
  at each fraction. Increasing $F_{\rm DM}$ instead shortens
  the canonical DM-led lifetime by several hundredfold between
  $1\%$ and $20\%$ for the complete coupling sequences,
  consistent with the approximate inverse-square trend
  discussed in \cref{sec:core_results}. Even at $20\%$,
  the resolved canonical DM-led modes remain approximately
  $8$--$15$ times longer-lived than their NM-led counterparts.

  Stronger self-repulsion expands the dark core and lowers
  its mode frequency, while also changing the relative fluid
  motion and metric response. The leading quadrupole
  diagnostic, $(\tau^{\rm quad})^{-1}\propto
  \omega^4|Q_{\rm N}+Q_{\rm D}|^2/J$, illustrates why a lower
  frequency favors weaker radiation at fixed normalized
  quadrupole strength; that strength, however, is not fixed
  across the scan. The full-GR damping time is determined
  by the complete coupled eigenfunction and its outgoing-wave
  matching, rather than by the frequency factor alone.
  The high-mass upturns, especially pronounced at
  $F_{\rm DM}=1\%$, further illustrate this distinction:
  the radiative lifetime can increase sharply while the
  frequency changes only modestly.

%%%%%%%%%%%%%%%%%%%%%%%%%%%%%%%%%%%%%%%%%%%%%%%

\subsection{Higher-multipole (\texorpdfstring{$\ell=3$}{ell=3}) modes}

\begin{figure*}[!tb]
    \centering
    \includegraphics[width=0.65\linewidth]{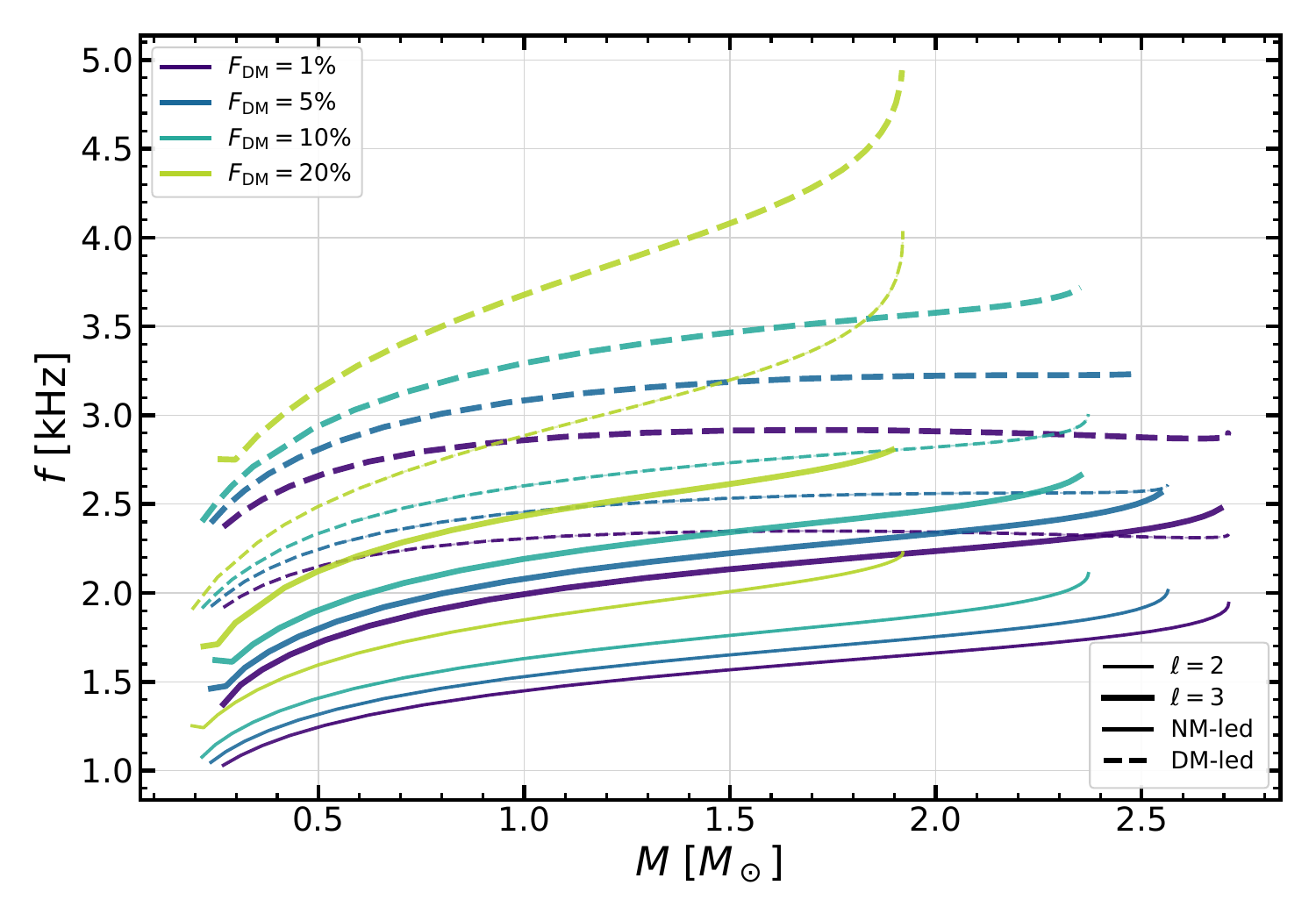}
    \caption{Full-GR quadrupolar ($\ell=2$) and octupolar
    ($\ell=3$) $f$-mode frequencies versus total gravitational
    mass for EOS1 with bosonic dark cores
    ($m_\chi=500\,\mathrm{MeV}$, $\lambda=\pi$).
    Thin and thick curves denote $\ell=2$ and $\ell=3$,
    respectively; solid and dashed curves denote the NM-led
    and DM-led branches. Colors distinguish
    $F_{\rm DM}=1\%,\,5\%,\,10\%$, and $20\%$.}
    \label{higher_l_modes}
  \end{figure*}

\cref{higher_l_modes} shows that the NM-led and DM-led
  families persist at $\ell=3$, with both branches lying above
  their quadrupolar counterparts. Increasing $\ell$ changes the
  angular restoring response of the coupled fluid--metric
  perturbations without changing the equilibrium star.
  At fixed mass, both octupolar frequencies increase with
  $F_{\rm DM}$, continuing the core trend of \cref{freq_mass}.
  Their multipolar spacing responds differently, however:
  at $1.4\,M_\odot$, increasing $F_{\rm DM}$ from $1\%$ to
  $20\%$ decreases $f_3/f_2$ from $1.36$ to $1.30$ for the
  NM-led branch, but increases it from $1.24$ to $1.28$ for
  the DM-led branch. The two families therefore cannot be
  related across multipoles by a common, fraction-independent
  factor. Both ratios lie below the Kelvin value
  $\sqrt{15/7}\simeq1.46$ for a homogeneous, incompressible
  Newtonian fluid sphere; that limiting result does not capture
  the nonuniform, compressible two-fluid structure considered
  here. The NM-led $\ell=3$ and DM-led $\ell=2$ frequency
  bands overlap. Since different $\ell$ sectors decouple in
  the spherical, nonrotating background, such frequency
  coincidences do not imply mode mixing. They nevertheless
  make a single observed frequency insufficient to distinguish
  an NM-led octupole from a DM-led quadrupole without additional
  information about the star or the oscillation.

% --- ORIGINAL BEGIN ACTIVE_BIBLIOGRAPHY_FILE
\iffalse
\bibliography{refrences}
\fi
% --- ORIGINAL END ACTIVE_BIBLIOGRAPHY_FILE
\bibliography{refrences_active}

\end{document}